\documentclass[
    reprint,
    prd,
    nofootinbib,
    amsmath,
    amssymb,
    aps,
    floatfix,
]{revtex4-2}

\usepackage[T1]{fontenc}
\usepackage{aas_macros}
\usepackage{bm}
\usepackage{graphicx}
\usepackage[dvipsnames]{xcolor}
\usepackage{hyperref}

\begin{document}

\title{Conic Tilings of Black Hole Images by Radial Closest Approach}

\author{George N.~\surname{Wong}\textsuperscript{$+$}}
\email[E-mail: ]{gnwong@ias.edu}
\affiliation{School of Natural Sciences, Institute for Advanced Study, 1 Einstein Drive, Princeton, NJ 08540, USA}
\affiliation{Leinweber Forum for Theoretical Physics, Princeton University, Princeton NJ 08544, USA}

\author{Andrew \surname{Mummery}\textsuperscript{$+$}}
\email[E-mail: ]{amummery@ias.edu}
\affiliation{School of Natural Sciences, Institute for Advanced Study, 1 Einstein Drive, Princeton, NJ 08540, USA}

\begingroup
\renewcommand{\thefootnote}{$+$}
\footnotetext{These authors contributed equally to this work.}
\endgroup

\date{\today}

\begin{abstract}
Each point in a black hole image corresponds to a light ray arriving at the observer along a particular direction. When traced backward from the observer, some rays travel inward, reach a minimum radius, and then turn back outward. Grouping these rays by their minimum radius reveals a simple geometric structure in the image. In the standard Bardeen image coordinates for a distant Kerr observer, these contours are arcs of conic sections, i.e., segments of ellipses, parabolas, and hyperbolas. We show that this conic structure is a feature of natural local image coordinates at finite radius not only in Kerr but in the broader Pleba\'{n}ski--Demia\'{n}ski (PD) family of exact Einstein--Maxwell solutions containing Kerr, with observer motion accounted for by aberration. To find these contours, we identify the image points whose rays would have zero radial velocity at a chosen radius. These points lie on a complete conic, from which we remove points whose rays turn back farther out. In Kerr, we determine these arcs explicitly and show how they tile the returning image as the minimum radius varies. In PD, for rays with positive Killing energy, we construct these contours and prove that the field equations and alignment of the principal null directions impose a coefficient identity that makes a local test of the radial potential's slope sufficient to identify the first inward radial root. This result holds throughout every connected positive-radius region where the radial metric function $\Delta_r$ is positive and radial motion is allowed at both endpoints, assuming that the spacetime mass parameter is positive. We then explicitly show how this local test can fail in a radial domain that crosses $r = 0$ and in the separable Ovcharenko--Podolsk\'{y} extension of PD, where a ray can turn back farther out instead of approaching the unstable photon orbit that passes the local test. In Kerr and PD, the conic family also gives a single polynomial equation that the critical curve must satisfy, restricting its shape for a given spacetime and observer. These results recast the returning image as a geometric, topographic map of radial photon propagation, showing how deeply returning rays probe the spacetime and constraining where their light can originate.
\end{abstract}

\maketitle

\section{Introduction}
\label{sec:introduction}

The electromagnetic appearance of a black hole depends on both its astrophysical environment and the geometry of spacetime, which determines how light propagates. Each point on an image selects a direction on the sky and therefore corresponds to a particular null geodesic through the spacetime. The emissivity, absorption, and dynamics of the source determine the light accumulated along the ray, but the spacetime and observer determine the path of the ray itself. Null geodesics near black holes have therefore been studied since the earliest calculations of escape cones and the optical appearance of compact objects and now play a central role in interpreting horizon-scale images \cite{synge_1996_escapephotons,cunningham_1973_starkerrbh,luminet_1979_blackholeimage,falcke_2000_shadow,perlick_2004_gravlensing,eht_m87_1,eht_m87_5,eht_m87_6,psaltis_2020_ehtgrshadow,eht_sgra_1,eht_sgra_5,eht_sgra_6}.

\begin{figure*}[t]
    \centering
    \includegraphics[width=.9\linewidth]{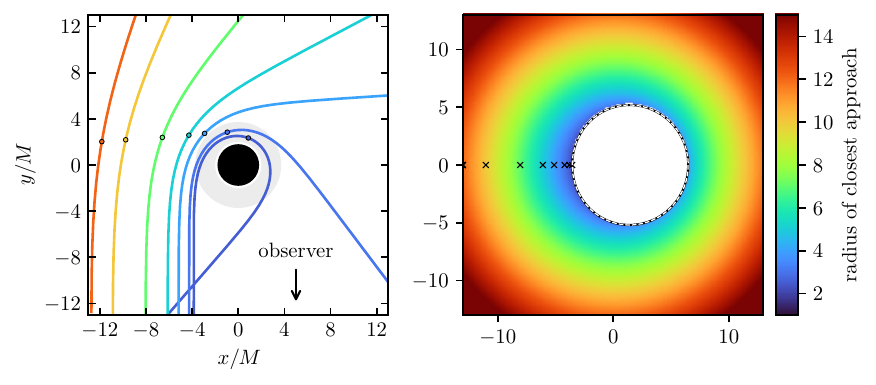}
    \caption{Closest approach of rays traced backward from a distant equatorial observer in Kerr with $a/M = 0.7$. The left panel follows a set of equatorial rays, with black points marking their radial turning points. The black disk denotes the event horizon, and the gray annulus spans the prograde-to-retrograde equatorial photon-orbit radii. The right panel maps the radius of closest approach across the observer's image. Color encodes the turning radius $R/M$ in both panels, and black crosses identify the image points associated with the rays shown on the left. The white region contains captured rays with no exterior radial turn, and its dashed boundary is the critical curve. As an image point approaches this boundary from outside, its ray passes increasingly close to an unstable photon orbit and undergoes stronger deflection.}
    \label{fig:kerr-geodesics}
\end{figure*}

In this paper, we organize light rays by the radius at which they turn back outward when traced inward from the observer. These turning radii define a map of closest approach that is independent of the source of emission. For any chosen radius, we first identify the image points whose rays have a possible radial turn there. These points lie on a conic in image coordinates built from the photon directions measured by the observer, and varying the chosen radius generates a family of conics across the image. An inward ray can encounter a larger turning point before reaching the chosen radius, however, so we retain only the points whose rays actually turn there to form the physical closest-approach contour. Figure~\ref{fig:kerr-geodesics} illustrates this assignment for equatorial Kerr rays, with each trajectory colored by its turning radius. To generalize, we ask how this conic tiling construction changes when the black hole is charged or accelerating, or when the cosmological constant is nonzero, and study these effects within the Pleba\'{n}ski--Demia\'{n}ski (PD) family of spacetimes, which includes Kerr \cite{plebanski_1976_pdfamily}. In both Kerr and PD, a single projection of the photon directions measured by the observer preserves the full family of conics across all turning radii.

To select the image points whose geodesics reach a given minimum radius, we ask two questions:
\begin{enumerate}
    \item Which image points correspond to rays whose radial potential has a root at $R$, indicating a possible turn there?
    \item Which of those rays actually turn at $R$ when traced inward from the observer?
\end{enumerate}
The first question is local, since it uses only the radial potential at $R$, whereas the second question generally requires a global comparison of the accessible roots. Nevertheless, we show when the sign of the radial potential's derivative at $R$ can select the physical arcs directly, obviating the need to perform a global comparison.

Perhaps the most familiar consequence of gravity's effect on light propagating near a black hole is the black hole shadow, whose boundary is often called the critical curve. The portion reached by initially inward rays bounds the closest-approach map, since these rays asymptotically approach unstable spherical photon orbits rather than undergoing a finite radial turn \cite{bardeen_1973_kerrgeo,teo_2003_photonskerr,grenzebach_2014_kerrnewmannutshadows,perlick_2022_analyticbhshadows}. Once the spacetime and observer are specified, the critical curve and closest-approach map are fixed, although the intensity along each ray depends on the messier gastrophysics, including the distribution, dynamics, and radiative properties of the emitting plasma \cite{gralla_2019_bhshadows,kocherlakota_2022_emissionvsgravity,ozel_2022_bhgrtests,younsi_2023_bhgrtests,eht_m87_5,eht_sgra_5}.

For an emitting disk, previous work has used transfer functions and lensing-band calculations to relate image points to the radii at which their rays cross the disk \cite{cunningham_1975_transferfunc,rauch_1994_caustics,dexter_2009_geokerr,gralla_2020_kerrlensing,cardenasavendano_2023_aart,cardenasavendano_2024_lensingbands}. Our closest-approach radius records how deeply a ray travels, independently of where the light is emitted. Closest-approach contours are circles in spherical symmetry \cite{kocherlakota_2022_emissionvsgravity,kocherlakota_2024_hotspots}, and related conics appear in studies of wormholes and compact objects with material surfaces, where one asks which rays reach the throat or surface \cite{shaikh_2018_rotatingwormholeshadow,gyulchev_2018_wormholeshadow,shaikh_2021_kerralternatives,wang_2023_horizonornotshadows}. The throat or surface sets a particular radius in those examples, whereas here we follow closest-approach contours across the full range of turning radii. Together, these contours provide a common reference for comparing emission models by identifying which parts of the image probe small radii.

We first examine how these contours change with radius in Kerr. As the candidate turning radius decreases, the complete conics pass from ellipses through a parabola to hyperbolas. We derive these curves using Carter's conserved quantity, which lets us separate the equations for a light ray's radial and angular motion \cite{carter_1968_hamiltonjacobi,carter_1968_globalstructurekerr,walker_1970_penrosewalker}. We also show that the slope of the radial potential at a proposed turn tells us which rays actually turn there, allowing us to directly select the physical closest-approach contours.

We then ask how much of this construction survives in the Pleba\'{n}ski--Demia\'{n}ski family of Einstein--Maxwell solutions with a cosmological constant. The PD family contains Kerr and its charged generalization, Kerr--Newman, but also allows acceleration, cosmological curvature, and gravitomagnetic (NUT) charge \cite{plebanski_1976_pdfamily}. These extensions preserve the separation of radial and angular photon motion, and every ray's radial equation contains a shared function whose coefficients are fixed by the spacetime \cite{kubiznak_2007_killingyanopd,krtous_2008_generalhiddensymmetries}. Grenzebach, Perlick, and L\"{a}mmerzahl (GPL) used this separation property to derive the shadow boundary from spherical photon orbits \cite{grenzebach_2014_kerrnewmannutshadows,grenzebach_2015_gplpd}. We use the same ray equations to construct closest-approach contours for returning rays that have positive Killing energy.

Even though the PD family preserves separability, radial motion in PD can be more complicated than in Kerr, so identifying where an inward ray actually turns might appear to require a global comparison. We show that, as long as the PD mass is positive, the structure of the photon radial potential and a coefficient identity imposed by the field equations again reduce this comparison to a local test of the potential's slope. Our result holds on every connected positive-radius interval where the shared radial function is positive (with radial motion allowed at the inner boundary and at the observer). It also applies when the interval extends to infinity, provided radial motion is allowed at sufficiently large radii. When NUT charge vanishes, we show that it is possible to further classify the possible radial turns in the allowed outer regions. In this case, each ray's radial potential then has at most two distinct roots, with the outer root giving the closest approach of an inward returning ray. We also construct a simple counterexample in a radial interval crossing $r=0$ to show how the result breaks down when the positive-radius restriction is lifted.

The PD solutions also have a special geometric alignment: tidal gravity and the electromagnetic field select the same pair of preferred lightlike directions, called principal null directions.\footnote{Along the gravitational principal null directions, the shearing effects of tidal gravity are symmetric such that a circular light beam is not sheared into an ellipse. Similarly, a (non-null) electromagnetic field has two principal null directions: in a local frame where the electric and magnetic fields are parallel or one vanishes, these point in opposite senses along the field axis.} For a spherical charged black hole, the gravitational pair consists of the ingoing and outgoing radial lightlike directions. The electromagnetic pair consists of the directions along and opposite to the radial electric field, so the two pairs coincide. This alignment persists in the rotating PD solutions. In contrast, the recent solutions found by Ovcharenko and Podolsk\'{y} (OP) allow the two fields to select different pairs while preserving the separation needed for the conic construction \cite{ovcharenko_2025_nonaligned,gray_2026_hidden}. We use this broader OP family to construct an exact counterexample for our tiling result: the local radial conditions identify an unstable fixed-radius photon orbit, yet the inward ray from the corresponding image point turns at a larger radius before it can approach that orbit. The candidate conics therefore survive when alignment is relaxed, even though an outer turn can hide an unstable orbit at positive radius. Thus, the PD coefficient identity supplies the additional restriction that excludes this ordering and makes the local radial test sufficient.

In Kerr and in PD, the underlying conic structure also explains the geometric restrictions on the shape of the critical curve. Combining the radial conditions for fixed-radius photon orbits lets us eliminate the orbit radius and describe their candidate image points with a single implicit relation whose coefficients are fixed by the spacetime and observer. In adapted image coordinates, this recovers the known sixth-degree Kerr polynomial \cite{gralla_2020_observableshape} and gives a reflection-symmetric polynomial of degree eight for generic PD. More broadly, the closest-approach contours together with their critical boundary form a topographic map of radial photon propagation connecting image shape to the depths probed by rays. In Kerr, for example, the flattened prograde side of the critical curve is associated with rays that probe smaller radii than those on the rounded retrograde side. The map provides a common way to compare image geometry across spacetimes independently of the emission model.

The remainder of this paper is organized as follows. In Section~\ref{sec:general-conics}, we overview the conic construction and the conditions that select physical image curves. In Section~\ref{sec:kerr}, we provide a full example of the construction in Kerr. Section~\ref{sec:pd} extends the construction to PD and examines cases where the local radial test fails, both in a domain crossing $r = 0$ and in the OP family. Finally, in Section~\ref{sec:discussion}, we summarize our results and discuss their implications for image geometries. The appendices collect proof details, examples, screen limits, and Kerr extensions.

\section{Closest-approach geometry on the observer screen}
\label{sec:general-conics}

\subsection{The closest-approach map}
\label{sec:closest-approach-map}

The right panel of Figure~\ref{fig:kerr-geodesics} is effectively a topographic map of radial photon propagation. Each colored image point corresponds to a light ray traced backward from the observer, with its color indicating the smallest radial coordinate $R$ reached by the ray. Rays with large $R$ turn far from the black hole, while rays with small $R$ probe more deeply. Curves of constant $R$ form contours on this map, the white region contains rays with no finite radial turn, and its dashed boundary is the critical curve.

In a separable spacetime, each ray's radial motion can be expressed in terms of a radial potential $\mathcal{R}(r)$.\footnote{We follow the standard Kerr terminology in referring to the radial function $\mathcal{R}$ as a potential.} Up to a positive metric factor and a choice of ray parameter $s$,
\begin{align}
    \left(\frac{{\rm d}r}{{\rm d}s}\right)^2 \propto \mathcal{R}(r).
\end{align}
Radial motion is allowed where $\mathcal{R} \geq 0$. For a returning ray (traced inward from the observer), the first root it reaches determines its physical turning radius, because radial motion is forbidden immediately inside that radius. At a regular point on the critical curve, this outer root has multiplicity two, and the ray approaches the corresponding unstable photon orbit asymptotically, rather than turning around.

The radial potential could also have roots at smaller radii that the light ray cannot reach, which complicates the question of identifying the physical minimum radius: the $\mathcal{R}$ for an observer at radius $r_o$ could have two simple roots at $R_1 < R_2 < r_o$. In that case, then because the ray reaches $R_2$, it turns before it can reach $R_1$, and although the image point formally belongs to the fixed-root curves for both $R_1$ and $R_2$, only the $R_2$ curve describes its physical closest approach. The difference between having a root at a given radius and actually reaching that radius is thus central to the construction of the map.

The directions of arriving photons on the local sky depend on the observer's position and motion. The image map is constructed by choosing a viewing orientation and projecting these directions onto a plane, and changing this projection moves image points while preserving each ray's turning radius. Although most black hole images show a small part of the sky facing the hole, the image for a general observer may require describing the full sky, which also includes the outward view. For the stationary observers used below, the same transverse image coordinates can describe a ray in either view, so we specify whether backward tracing initially moves toward smaller or larger $r$. The two views meet along directions where the ray initially has no radial motion.

\begin{figure*}[t]
    \centering
    \includegraphics[width=\linewidth]{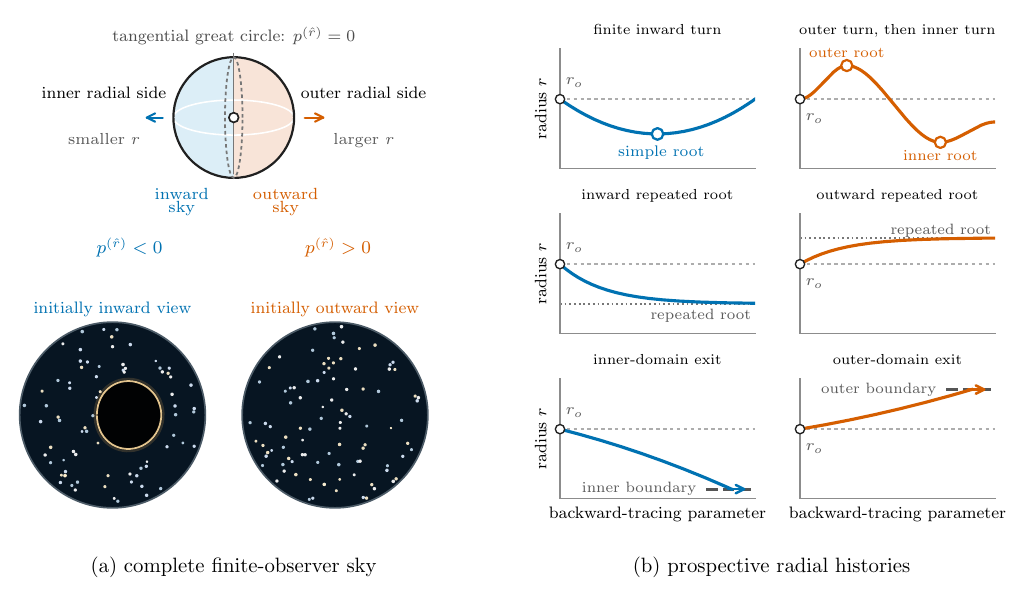}
    \caption{Possible radial histories of rays traced backward from a stationary observer at finite radius. On the left, directions with no initial radial motion form the tangential great circle $p^{(\hat r)} = 0$, which separates the inward and outward views. The arrows indicate the directions of decreasing and increasing $r$. The two lower images show a schematic Kerr example at large observer radius: the inward view contains the ``black hole shadow'' capture region, while the outward view looks toward the surrounding sky. On the right, blue and orange curves follow rays that initially move inward and outward, respectively. A ray can reach a simple radial root and turn, approach a repeated root asymptotically, or leave through a boundary of the observer's radial domain. An initially outward ray can also turn back inward, pass below the observer radius, and reach an inner turn. The closest-approach construction that we consider in this work focuses on initially inward rays that turn back outward.}
    \label{fig:radial-histories}
\end{figure*}

Figure~\ref{fig:radial-histories} categorizes the different radial histories of rays that are (initially) traced inward or outward from the observer. An inward ray can turn at a finite simple root, approach a finite repeated root asymptotically, or leave the observer's radial domain through its inner boundary. Similarly, an outward ray can continue to the outer boundary, approach a repeated root outside the observer, or reach a maximum radius, turn inward, and pass back below $r_o$. After turning inward, the ray can turn again at a smaller radius or leave through the inner boundary. Depending on the asymptotic geometry, the limiting ray between return and escape can also approach infinity with vanishing radial motion. These possibilities are largely hidden by the familiar distant Kerr camera.

For observers at finite radius in the PD family, the radial domain can include outer horizons, conformal boundaries, and barriers that return an initially outward ray. Describing the complete sky therefore requires keeping track of each ray's initial radial direction and its subsequent radial motion. Within this broader picture, our conic tiling assigns closest-approach radii to rays that initially travel inward and then return, with the largest accessible root below the observer giving the minimum radial coordinate reached. Throughout this paper, we measure radial depth using the coordinate $r$ in the separated geodesic equations, although note that this coordinate does not necessarily measure proper distance.\footnote{In Kerr and the nonaccelerating geometries with a principal tensor, $r$ is geometrically distinguished up to sign and branch choices as an eigenvalue coordinate of the principal tensor \cite{krtous_2008_generalhiddensymmetries,frolov_2017_hiddensymmetries}. In Kerr, it can also be reconstructed locally from curvature invariants after fixing the asymptotic branch and normalization \cite{abdelqader_2015_kerrinvariants}. Separability by itself determines the radial foliation only up to a reparameterization $r\mapsto f(r)$. For the full PD family, we use the canonical radial coordinate and normalization of the GPL form \cite{kubiznak_2007_killingyanopd,grenzebach_2015_gplpd}.}

Let $(\alpha,\beta)$ denote (thus far arbitrary) coordinates on the inward-facing image, and define the closest-approach map by
\begin{align}
    T(\alpha,\beta) = \max\left\{r<r_o:\mathcal{R}(r;\alpha,\beta)=0\right\}, \label{eq:turning-map}
\end{align}
where we take the maximum over all roots accessible to the inward ray, with $T$ defined whenever such a root exists. At a regular point of the critical boundary, the ray approaches the radius $T$ asymptotically. Rays that leave through the inner boundary without turning have no finite turning-radius label, so $T$ is undefined at those image points.

Image points with the same closest-approach radius then trace out the physical contour
\begin{align}
    \mathcal{T}_R =\left\{(\alpha,\beta):T(\alpha,\beta)=R\right\}. \label{eq:physical-radius-contour}
\end{align}
To construct this contour, we first collect all image points whose radial potential has a root at $R$:
\begin{align}
    \mathcal{Z}_R =\left\{(\alpha,\beta):\mathcal{R}(R;\alpha,\beta)=0\right\}. \label{eq:prescribed-root-locus}
\end{align}
Every point of $\mathcal{T}_R$ belongs to $\mathcal{Z}_R$, but some points along the candidate curve may correspond to light rays that encounter a larger root first or fail the angular or observer-domain conditions needed to reach $R$. The radial equation at $R$ therefore determines $\mathcal{Z}_R$ locally, while the ray's global radial and angular motion selects the physical contour $\mathcal{T}_R$.

\subsection{Fixed-radius conics}
\label{sec:general-conic-reduction}

The curve $\mathcal{Z}_R$ collects all image points whose radial potentials have a root at $R$, and its conic form follows from how the constants of motion enter the separated radial equation. As we show below for Kerr and PD, the radial potential is quadratic in the axial impact parameter and linear in the remaining separation constant. We choose image coordinates in which the axial impact parameter is linear up to a constant offset and the separation constant is quadratic. The familiar horizontal and vertical impact coordinates used for distant Kerr images are one such physically motivated choice. Substituting these expressions at a fixed radius therefore gives a quadratic equation on the image.

A photon moving through a stationary, axisymmetric spacetime conserves its Killing energy $E = -p_t$ and axial angular momentum $L_z = p_\phi$. Its four-momentum satisfies the null condition $g^{\mu\nu} p_\mu p_\nu = 0$, which is quadratic in the momenta. When this condition separates into a sum of radial and angular terms, the radial equation remains quadratic in $E$ and $L_z$, with the separation constant $K$ entering linearly \cite{carter_1968_hamiltonjacobi,gray_2026_hidden}. For $E \neq 0$, dividing the radial equation by $E^2$ then gives the normalized potential
\begin{align}
    \mathcal{R}(r;\xi,\lambda) =A(r)+B(r)\xi+C(r)\xi^2+D(r)\lambda, \label{eq:general-radial-form}
\end{align}
where $\xi = L_z/E$ is the axial impact parameter and $\lambda = K/E^2$ is the normalized separation constant. To obtain conics on the image, it is also required that the image coordinates satisfy
\begin{align}
    \xi&=p_0+p_1\alpha, \nonumber \\
    \lambda&=q_0+q_1\alpha+q_2\alpha^2+q_3\beta^2. \label{eq:general-screen-map}
\end{align}
The coefficients $\{p_i, q_i\}$ are fixed by the metric, the observer's position and local frame, and the normalization of the image coordinates. These coefficients are the same for every image point and trial radius, so one coordinate system describes the entire family. The radial and image-coordinate relations above are sufficient for the conic construction and allow arbitrary radial dependence in $A,B,C,D$.

At a chosen radius $R$, substituting these expressions into $\mathcal{R}(R)=0$ gives
\begin{align}
    \mathcal{C}_0(\alpha,\beta;R) &=c_0(R)+\ell_0(R)\alpha+a_0(R)\alpha^2 +b_0(R)\beta^2=0, \label{eq:general-root-conic}
\end{align}
with
\begin{align}
    c_0& = A_R + B_R p_0 + C_R p_0^2+D_R q_0, \nonumber \\
    \ell_0& = B_R p_1+2 C_R p_0 p_1+D_R q_1, \nonumber \\
    a_0& = C_R p_1^2 + D_R q_2, \nonumber \\
    b_0 &= D_R q_3. \label{eq:general-conic-coefficients}
\end{align}
where $A_R = A(R)$ and likewise for $B_R$, $C_R$, and $D_R$.

Equation~\ref{eq:general-root-conic} thus shows that, for any radius $R$, all candidate image points in $\mathcal{Z}_R$ lie on one (possibly empty or degenerate) conic. Varying $R$ changes the coefficients and generates the full family of candidate conics. For a nondegenerate conic, $a_0 b_0 > 0$ gives an ellipse or no real image points, $a_0 b_0 < 0$ gives a hyperbola, and $a_0 = 0$ with $b_0 \ell_0 \neq 0$ gives a parabola. The conic form is preserved under affine or projective changes of image coordinates, which can introduce mixed and additional linear terms while preserving its degeneracy class.

To select the physical arcs, we first examine the potential near a candidate root. Since radial motion is allowed where $\mathcal{R} \geq 0$, its derivative at a simple root tells us whether the ray can propagate immediately outside that radius. We take this derivative at fixed constants of motion (equivalently at fixed image coordinates $(\alpha, \beta)$) and define
\begin{align}
    \mathcal{C}_1(\alpha,\beta;R) =\partial_r\mathcal{R}(R;\alpha,\beta), \label{eq:general-derivative-conic}
\end{align}
Differentiation with respect to $r$ preserves the quadratic form in the image coordinates, so $\mathcal{C}_1=0$ defines a second conic. Its coefficients follow from Equation~\ref{eq:general-conic-coefficients} by replacing $A, B, C, D$ with their radial derivatives. The sign of $\mathcal{C}_1$ enters the selection of physical arcs. At intersections with $\mathcal{C}_0 = 0$, the radial potential and its first derivative both vanish, so these image points identify repeated-root candidates. Appendix~\ref{app:general-conic-details} gives a criterion for conic degeneracy and conditions under which these repeated-root image points vary smoothly with $R$.

\subsection{Selecting the physical arcs}
\label{sec:general-branch-selection}

At a point on the candidate conic, the ray's radial potential has a root at $R$ by definition, but for $R$ to give the physical closest approach, the inward ray must reach it before encountering a larger root. At a simple turning point reached from larger $r$, radial motion is allowed immediately outside $R$ and forbidden immediately inside it, so
\begin{align}
    \mathcal{R}(R)=0, \qquad \mathcal{R}'(R)>0.
\end{align}
where a prime denotes differentiation with respect to $r$ at a fixed image position. A positive slope is necessary for such a turn, but a deeper root blocked by a larger one could in principle have the same local behavior. In general, selecting the physical root therefore requires comparing the real roots below the observer in the radial domain of interest.

For each image point, we discard roots that the ray cannot reach within its allowed radial and angular motion and retain the largest remaining root. This selects the points of each candidate conic that belong to the physical contour:
\begin{align}
    \mathcal T_R =\mathcal Z_R\cap\{(\alpha, \beta) : R = T(\alpha, \beta)\}. \label{eq:physical-arc-selection}
\end{align}

We can visualize this selection by stacking the candidate conics above the image plane, placing each conic at a height $R$. At each image point, the highest accessible intersection below $r_o$ gives its closest-approach radius $T(\alpha,\beta)$, and keeping this intersection at each image point traces a surface whose height records closest approach, with gaps wherever rays have no accessible root. Horizontal slices through this surface give the physical contours $\mathcal{T}_R$.

One of our main results is that this global radial selection reduces to a local test in Kerr and PD, under the general conditions specified below. A positive slope at a candidate root then guarantees that no larger root blocks the inward ray from reaching it. In PD, this guarantee follows from a coefficient identity imposed by the field equations that restricts how the radial roots can be ordered. We can therefore construct the closest-approach surface one contour at a time using implicit equations on the image. Each conic equation defines the candidate curve, and the local derivative-sign condition selects its physical arcs. Varying $R$ then builds the tiling without ever requiring a comparison between the different radial roots at every image point.

\subsection{The critical curve as a limiting boundary}
\label{sec:critical-fold}

In the positive-radius Kerr and PD black hole exteriors we consider below, we can use the same conic construction to algebraically construct the portion of the critical curve that bounds the initially inward returning image.\footnote{For a finite observer, this portion has $R < r_o$. Critical rays with $R > r_o$ initially travel outward and belong to the outward view. Identifying the inward boundary with the full critical curve requires the observer to lie outside all relevant visible photon-orbit radii.} As an image point approaches this boundary, its ray's turning point tends toward an unstable constant-radius photon orbit, which the limiting ray approaches asymptotically. At a generic point on the boundary, the selected radial root satisfies
\begin{align}
    \mathcal{C}_0&=\mathcal{C}_1=0, & \mathcal{C}_2&\equiv\partial_r^2\mathcal{R}>0, \label{eq:physical-fold-conditions}
\end{align}
where the positive second derivative fixes the repeated root's multiplicity to two. On the image, the candidate points lie where the root and derivative conics intersect, so tracing the intersection between the conics and varying $R$ traces the candidate critical curve.

In a general separable spacetime, a larger radial turn can hide an unstable photon orbit, placing its candidate image point inside the returning image. The initially inward portion of the physical critical curve is traced by limiting rays that approach the orbit from larger radius along an angularly allowed, regular path within the observer's domain. In the Kerr and PD black hole exteriors considered below, these rays separate return from capture. More generally, they can separate returning rays with different turning radii.\footnote{Appendix~\ref{app:general-conic-details} gives a condition for the conic intersections to vary smoothly with $R$ and explains the special treatment needed on the image's reflection axis. Roots of higher multiplicity, conics that are tangent or share a component, rays initially tangent to a surface of constant $r$, and roots on a radial boundary require separate local treatment.} Under the conditions we provide below for Kerr and PD, the local radial test rules out this obstruction. In Kerr and PD, eliminating $R$ gives a single polynomial equation for the candidate critical curve, with coefficients fixed by the spacetime and observer. The same conic family therefore determines both the closest-approach contours and the algebraic curve containing their critical boundary.

\subsection{Contour crowding near the critical curve}
\label{sec:contour-spacing}

Our conic construction also provides a direct way to compute how the closest-approach contours crowd together near the critical curve. As their spacing shrinks, the radial depth probed by the rays changes rapidly across the image. To calculate this spacing, we consider a smooth family of returning rays with simple turning roots, whose closest-approach radii satisfy
\begin{align}
    \mathcal{C}_0\bigl(\alpha,\beta;T(\alpha,\beta)\bigr)=0,
\end{align}
and differentiating with respect to the image coordinates gives
\begin{align}
    \nabla_{\rm scr} T =-\frac{\nabla_{\rm scr}\mathcal{C}_0}{\mathcal{C}_1}, \label{eq:contour-gradient}
\end{align}
where $\nabla_{\rm scr}$ denotes the gradient with respect to the image coordinates $(\alpha,\beta)$ on the observer screen. On the right-hand side, we first differentiate $\mathcal{C}_0$ at fixed $R$, then substitute $R=T(\alpha,\beta)$ into the resulting gradient and the denominator $\mathcal{C}_1$. Where $\nabla_{\rm scr}\mathcal{C}_0 \neq 0$, let $n$ denote (signed) distance normal to a contour, increasing along this gradient. The spacing per unit change in $R$ is then
\begin{align}
    \frac{{\rm d}n}{{\rm d}R} =-\frac{\mathcal{C}_1} {\lVert\nabla_{\rm scr}\mathcal{C}_0\rVert}. \label{eq:contour-spacing}
\end{align}
As an image point approaches a regular point of the critical curve from the returning side, $\mathcal{C}_1$ tends to zero. If $\lVert\nabla_{\rm scr}\mathcal{C}_0\rVert$ remains nonzero, Equation~\ref{eq:contour-spacing} shows that the image separation per unit change in turning radius also tends to zero, producing the contour crowding. Combined with an emission and radiative transfer model, this map of radial depth can help interpret narrow image features.

We now carry out the full construction in Kerr to show how the selected conic arcs tile the returning image. The later PD analysis will then show when the coefficient identity imposed by the field equations lets the local slope test select the first radial root reached by an inward ray.

\section{Closest approach in Kerr}
\label{sec:kerr}

In subextremal Kerr, the null geodesic equations separate, and the standard impact coordinates connect the constants of motion to a distant observer's image \cite{carter_1968_globalstructurekerr,chandrasekhar_1983_blackholes,gralla_2020_nullgeodesics}. We use these relations to construct the candidate conics, whose shapes change between ellipses, a parabola, and hyperbolas as the trial radius varies. We then show that a positive slope of the radial potential selects the points on each conic where inward rays actually turn. The limiting points where this slope vanishes trace the critical curve boundary of the black hole shadow.

For the distant observer, we take $(\alpha,\beta)$ to be Bardeen's horizontal and vertical impact coordinates \cite{bardeen_1973_kerrgeo}, obtained by scaling the angular offsets on the sky by the observer distance as $r_o \to \infty$. We later demonstrate how to extend this construction to the more general zero angular momentum observer (ZAMO) at finite radius. Extremal horizon components and general observer aberration are discussed in Appendix~\ref{app:kerr-extensions}.

\subsection{Radial potential and distant screen}

We use geometrized units $G = c = 1$ and a metric signature $(-, +, +, +)$. In Boyer--Lindquist coordinates $x^\mu = (t, r, \theta, \phi)$, the Kerr metric is
\begin{align}
    ds^2 &= -\left(1-\frac{2Mr}{\Sigma}\right)dt^2 -\frac{4Mar\sin^2\theta}{\Sigma}\,dt\,d\phi \nonumber \\
    &\quad +\frac{\Sigma}{\Delta}\,dr^2 +\Sigma\,d\theta^2 \nonumber \\
    &\quad +\left(r^2+a^2 +\frac{2Ma^2r\sin^2\theta}{\Sigma}\right) \sin^2\theta\,d\phi^2, \label{eq:kerr-metric}
\end{align}
where $M$ is the mass of the black hole, $a = J/M$ is its angular momentum per unit mass, and
\begin{align}
    \Sigma &= r^2+a^2\cos^2\theta, & \Delta &= r^2-2Mr+a^2.
\end{align}
The radial metric function $\Delta$ is the Kerr limit of $\Delta_r$ in the PD notation used below. The outer horizon is the largest root of $\Delta$, which for $0 \leq |a| < M$ lies at
\begin{align}
    r_+=M+\sqrt{M^2-a^2}.
\end{align}

For a photon with four-momentum $k^\mu$, stationarity and axial symmetry give the conserved Killing energy $E = -k_t$ and axial angular momentum $L_z = k_\phi$. Separation of the Hamilton--Jacobi equation ensures the existence of the separate Carter constant $Q$ \cite{carter_1968_hamiltonjacobi,walker_1970_penrosewalker}. For $E \neq 0$, we normalize these constants by the photon energy and define
\begin{align}
    \xi&=\frac{L_z}{E}, & \eta&=\frac{Q}{E^2}, & \kappa&=\eta+(\xi-a)^2. \label{eq:kerr-constants}
\end{align}
The combination $\kappa$ is the energy-normalized radial separation constant $K/E^2$, where $K = Q + (L_z - aE)^2$.

These constants determine the radial and polar motion through the independent potentials
\begin{align}
    \mathcal{R}(r) &= \left(r^2+a^2-a\xi\right)^2-\kappa\,\Delta, \label{eq:kerr-radial} \\
    \Theta(\theta) &= \eta+a^2\cos^2\theta-\xi^2\cot^2\theta. \label{eq:kerr-polar}
\end{align}
For an affine parameter $s$, the equations of motion are
\begin{align}
    \left(\Sigma\frac{{\rm d}r}{{\rm d}s}\right)^2
    &=E^2\mathcal{R}(r),
    &
    \left(\Sigma\frac{{\rm d}\theta}{{\rm d}s}\right)^2
    &=E^2\Theta(\theta),
    \label{eq:kerr-separated-motion}
\end{align}
where again, the ray can propagate only where $\mathcal{R} \geq 0$ and $\Theta \geq 0$. A returning ray reaches its minimum radius at a simple root, where $\mathcal{R}(R)=0$ and $\mathcal{R}'(R) \neq 0$, and radial motion is forbidden immediately inside that radius. As the image point approaches the critical curve from the returning side, this turning root tends to a root of multiplicity two, satisfying
\begin{align}
    \mathcal{R}(R)&=\mathcal{R}'(R)=0, & \mathcal{R}''(R)&\neq0. \label{eq:kerr-multiplicity-two}
\end{align}
For these limiting rays, $\mathcal{R}''(R)>0$, so radial motion is allowed immediately on either side of the root. Expanding about $R$ gives
\begin{align}
    \mathcal{R}(r) = \frac{1}{2} \mathcal{R}''(R)(r-R)^2 + \mathcal{O}\bigl((r-R)^3\bigr), \label{eq:kerr-multiplicity-two-expansion}
\end{align}
so Equation~\ref{eq:kerr-separated-motion} gives $|{\rm d}r/{\rm d}s| = \mathcal{O}(|r-R|)$ when $\Sigma$ is regular. A critical ray therefore approaches $R$ only asymptotically, while a photon placed exactly there follows an unstable constant-radius, or spherical, orbit.

To construct the conics on the distant image, we write the radial potential in terms of $\xi$ and $\eta$, which makes its quadratic dependence on $\xi$ and linear dependence on $\eta$ explicit:
\begin{align}
    \mathcal{R}(r) &=r^4+(a^2-\xi^2-\eta)r^2 \nonumber \\
    &\quad +2Mr\left[(a-\xi)^2+\eta\right] -a^2\eta. \label{eq:kerr-radial-expanded}
\end{align}

For a distant observer at inclination $\theta_o$ with respect to the rotation axis, with $\sin\theta_o \neq 0$,\footnote{For an observer on the rotation axis, the image remains regular despite the singular expression for $\alpha$. The face-on view is obtained by taking $\theta_o \to 0$ or $\pi$.} Bardeen's impact coordinates are
\begin{align}
    \alpha &= -\frac{\xi}{\sin\theta_o}, & \beta &= \pm\sqrt{\Theta(\theta_o)},
\end{align}
where the two signs of $\beta$ correspond to the two directions of polar motion at the observer. Inverting these relations gives
\begin{align}
    \xi&=-\alpha\sin\theta_o, \label{eq:kerr-screen-xi} \\
    \eta&=\beta^2+(\alpha^2-a^2)\cos^2\theta_o. \label{eq:kerr-screen-eta}
\end{align}
For the analysis below, it is useful to express $\kappa$ directly in image coordinates, which conveniently makes the fact that it is nonnegative explicitly clear:
\begin{align}
    \kappa =\beta^2+(\alpha+a\sin\theta_o)^2\geq0. \label{eq:kerr-kappa-screen}
\end{align}

\subsection{Fixed-radius conics}
\label{sec:kerr-conics}

\begin{figure*}[t]
    \centering
    \includegraphics[width=\linewidth]{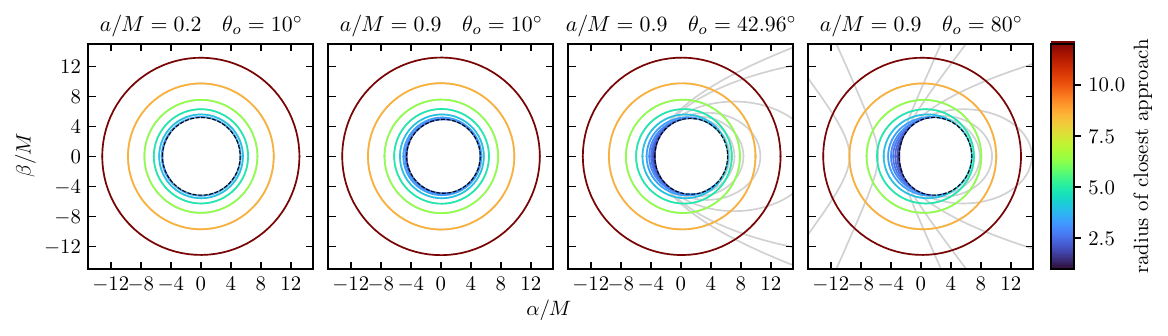}
    \caption{Closest-approach contours for a distant observer in Kerr. From left to right, the panels show $(a/M, \theta_o) = (0.2, 10^\circ)$, $(0.9, 10^\circ)$, $(0.9, \approx43^\circ)$, and $(0.9, 80^\circ)$. Gray curves show the complete candidate conics $\mathcal{R}(R) = 0$. The colored portions correspond to rays whose first radial turn is actually at $R$, with color giving $R/M$. The dashed black curve is the Kerr critical curve. The two nearly face-on views have similar tilings despite their different spins. At higher inclination, the conics become increasingly displaced and the small-radius contours concentrate on the prograde, negative-$\alpha$ side. The third panel lies near the onset of physical hyperbolic arcs, which appear along the prograde edge in the fourth panel.}
    \label{fig:kerr-tiling}
\end{figure*}

Each image point $(\alpha, \beta)$ determines the ray's constants $(\xi, \eta)$ and hence its radial potential. To find all candidate image points for a chosen radius $R > r_+$, we impose
\begin{align}
    \mathcal{R}(R;\alpha,\beta) = 0.
\end{align}
Equations~\ref{eq:kerr-screen-xi}--\ref{eq:kerr-screen-eta} express $\xi$ as a linear function of $\alpha$ and $\eta$ as a quadratic function of $(\alpha, \beta)$. Substituting these expressions into the expanded radial potential therefore gives the conic equation
\begin{align}
    \mathcal{C}_0(\alpha,\beta;R) &= C_0(R)+L_0(R)\alpha \nonumber \\
    &\quad +A_0(R)\alpha^2-B_0(R)\beta^2 = 0, \label{eq:kerr-root-conic}
\end{align}
where $\Delta_R=\Delta(R)$ and
\begin{align}
    B_0(R)&=\Delta_R, \nonumber \\
    A_0(R)&=-R^2+2MR-a^2\cos^2\theta_o, \nonumber \\
    L_0(R)&=4MaR\sin\theta_o,
\end{align}
and
\begin{align}
    C_0(R) &= R^4+a^2(1+\cos^2\theta_o)R^2 \nonumber \\
    &\quad +2MRa^2\sin^2\theta_o \nonumber \\
    &\quad +a^4\cos^2\theta_o.
\end{align}
Varying $R$ then generates the family of gray conics in Figure~\ref{fig:kerr-tiling}. The colored portions mark the image points whose rays reach the corresponding radius as their outer turn. Other points on the same conic can have a larger radial root that blocks the ray from reaching $R$.

To determine the shape of each conic, we can compare its horizontal and vertical quadratic terms. Outside of the horizon,
\begin{align}
    A_0(R)&=a^2\sin^2\theta_o-\Delta_R, & B_0(R)&=\Delta_R>0,
\end{align}
so the coefficient of $\beta^2$ remains negative while $A_0$ changes sign at
\begin{align}
    R_{\rm p} =M+\sqrt{M^2-a^2\cos^2\theta_o}. \label{eq:kerr-parabolic-radius}
\end{align}
For $a\sin\theta_o \neq 0$, decreasing $R$ through $R_{\rm p}$ changes the conic from an ellipse to a parabola and then a hyperbola. The metric in Equation~\ref{eq:kerr-metric} connects this transition to the outer ergoregion boundary:
\begin{align}
    A_0(R) = \Sigma(R,\theta_o)g_{tt}(R,\theta_o).
\end{align}
The outer ergoregion boundary satisfies $g_{tt} = 0$, so the parabolic radius $R_{\rm p}$ equals its radius at the observer's inclination $\theta_o$. For an equatorial observer, this places the parabolic member of the conic family at $R = 2M$ for any nonzero spin. The rays themselves can turn at a different polar angle, so their turning points need not lie on the ergoregion boundary.

For $A_0 \neq 0$, the linear term $L_0 \alpha$ shifts the conic horizontally, while the relative quadratic coefficients determine its eccentricity. Completing the square shows:
\begin{align}
    A_0(R)\bigl(\alpha-\alpha_0(R)\bigr)^2 -B_0(R)\beta^2+\widetilde C_0(R)=0, \label{eq:kerr-centered-conic}
\end{align}
where $\alpha_0$ is the horizontal position of the center and
\begin{align}
    \alpha_0(R) &= -\frac{L_0(R)}{2A_0(R)}, & \widetilde C_0(R)&=C_0(R)-\frac{L_0(R)^2}{4A_0(R)}.
\end{align}
As $R$ approaches $R_{\rm p}$, $A_0$ tends to zero and the center $\alpha_0$ recedes to infinity. The limiting curve is a parabola, which has no finite center. To check that the exterior conics never collapse to lines or a point, we evaluate
\begin{align}
    A_0(R)C_0(R) - \frac{L_0(R)^2}{4} =-\Delta_R\bigl(R^2 + a^2\cos^2\theta_o\bigr)^2<0. \label{eq:kerr-conic-nondegeneracy}
\end{align}
This expression is strictly negative outside the horizon, ensuring that every exterior conic is nondegenerate. At the parabolic transition, $A_0 = 0$, so the same relation requires $L_0 \neq 0$. In Schwarzschild, setting $a = 0$ removes the horizontal shift and gives $L_0 = 0$ and $A_0 = -B_0$, so every exterior candidate conic is a circle centered on the image origin as expected:
\begin{align}
    \alpha^2+\beta^2 = \frac{R^3}{R-2M}. \label{eq:schwarzschild-root-circles}
\end{align}

The selected contours include hyperbolic arcs only when returning rays can turn at $R < R_{\rm p}$. At the threshold, the parabolic conic touches the prograde edge of the critical curve at $\beta = 0$. Substituting $R = R_{\rm p}$ into the polar visibility condition in Equation~\ref{eq:kerr-critical-parametrization} gives the threshold inclination for observers between the rotation axis and the equatorial plane. For $1/\sqrt{2} \leq |a|/M \leq 1$, we find
\begin{align}
    \theta_{\rm crit} = \arcsin\sqrt{\frac{1+\sqrt{17-16(a/M)^2}}{8(a/M)^2}}.
    \label{eq:kerr-hyperbolic-inclination}
\end{align}
Physical hyperbolic arcs appear for $\theta_o > \theta_{\rm crit}$. For $a/M = 0.9$, the threshold is $\theta_{\rm crit} \simeq 42.96^\circ$, as in the third panel of Figure~\ref{fig:kerr-tiling}. At $10^\circ$, the physical contours are ellipses, while at $80^\circ$ they include hyperbolic arcs. For $|a|/M < 1/\sqrt{2}$, the physical contours remain ellipses at every inclination.

\subsection{Selecting the outer root}
\label{sec:kerr-branch}

To obtain the closest-approach contours, we next select the image points whose inward rays reach $R$ before any larger radial root. For a ray traced inward from a distant observer, the largest radial root outside the horizon determines its closest approach. In Kerr, the map therefore becomes
\begin{align}
    T(\alpha,\beta) = \max\{r>r_+:\mathcal{R}(r;\alpha,\beta)=0\}, \label{eq:kerr-turning-map}
\end{align}
whenever an exterior root exists. To select the physical part of a candidate conic, we must determine whether $R$ is this outermost root. We now show that the sign of $\mathcal{R}'(R)$ is sufficient to make that selection.

To determine whether a larger root can block the ray, we solve $\mathcal{R}(r) = 0$ for $\kappa$. At fixed $\xi$, the root condition becomes $\kappa = F_\xi(r)$, where
\begin{align}
    P_\xi(r)&=r^2+a^2-a\xi, & F_\xi(r)&=\frac{P_\xi(r)^2}{\Delta(r)}. \label{eq:kerr-F}
\end{align}
The curve $F_\xi$ is independent of $\kappa$, and the radial potential becomes
\begin{align}
    \mathcal{R}(r) =\Delta(r)\bigl[F_\xi(r)-\kappa\bigr], \label{eq:kerr-radial-level-set}
\end{align}
with $\Delta > 0$ outside the horizon. Each ray has fixed values of $\xi$ and $\kappa$, so its radial roots are the intersections of $F_\xi(r)$ with a horizontal line at its value of $\kappa \geq 0$. Radial motion is allowed where $F_\xi(r) \geq \kappa$. An inward ray moves toward smaller $r$, so the rightmost intersection determines its closest approach.

\begin{figure*}[t]
    \centering
    \includegraphics[width=0.9\linewidth]{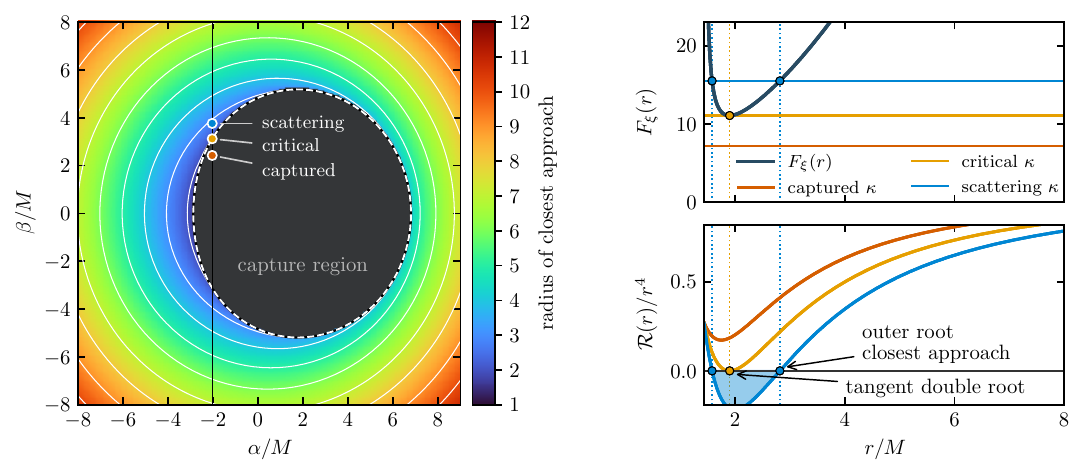}
    \caption{Capture, return, and approach to an unstable photon orbit in Kerr. The left panel shows the closest-approach map for $a/M = 0.9$ and $\theta_o = 80^\circ$, with color giving the turning radius $R/M$ of returning rays and white contours connecting points with the same closest approach. The dark gray region contains captured rays, which have no exterior radial turn, and the dashed black-and-white critical curve separates captured rays from returning rays. The vertical black line fixes $\xi$, and the three selected points identify the captured, critical, and scattering rays shown on the right. On the upper right, radial roots in the potential occur where $F_\xi(r)$ intersects the horizontal line at each ray's $\kappa$. For $\kappa$ below the minimum, the ray has no exterior root and is captured. At the minimum, the critical ray asymptotically approaches an unstable spherical photon orbit, where the potential has a root of multiplicity two with $\mathcal{R}''(R) > 0$. For $\kappa$ above the minimum, the ray has two exterior simple roots and reaches the larger one first as it travels inward. The lower-right panel shows the corresponding radial potentials. At the outer turn, the potential crosses from negative to positive as $r$ increases, so $\mathcal{R}'(R) > 0$. The ray never reaches the smaller root, where $\mathcal{R}'(R) < 0$.}
    \label{fig:kerr-radial}
\end{figure*}

Figure~\ref{fig:kerr-radial} illustrates this construction for rays along a vertical slice through the image. These rays share $\xi$ and therefore the same curve $F_\xi$, while their values of $\kappa$ set the horizontal levels. For the curve shown, a level below the minimum gives capture, a tangent level gives a critical ray, and a higher level gives an outer turn and return.

We now show that $F_\xi$ either decreases to one exterior minimum and then increases, or increases throughout the exterior region. In either case, an intersection on its increasing part must be the outermost root, which lets us directly associate the physical turn with the sign of $\mathcal{R}'$.

To determine where $F_\xi$ rises and falls, we factor its derivative:
\begin{align}
    F_\xi'(r) &=\frac{2P_\xi(r)G_\xi(r)}{\Delta(r)^2}, \nonumber \\
    G_\xi(r) &=2r\Delta(r)-(r-M)P_\xi(r). \label{eq:kerr-F-derivative}
\end{align}
The sign of this derivative is set by $P_\xi G_\xi$, since the denominator is positive. Because $P_\xi'(r) = 2r > 0$ in the exterior, $P_\xi$ has at most one exterior zero.

If $P_\xi$ vanishes at an exterior radius $r_{\rm z} > r_+$, then $F_\xi(r_{\rm z})=0$, and we need to show that this is its only exterior minimum. For $r_+ < r < r_{\rm z}$, $P_\xi < 0$ and
\begin{align}
    G_\xi(r) = 2r\Delta(r)-(r-M)P_\xi(r) > 0.
\end{align}
Thus $F_\xi$ decreases with increasing radius up to $r_{\rm z}$. Above $r_{\rm z}$, $P_\xi>0$, so $F_\xi$ increases as long as $G_\xi$ stays positive. At $r_{\rm z}$, $G_\xi(r_{\rm z})=2r_{\rm z}\Delta(r_{\rm z})>0$, and using $a\xi = r_{\rm z}^2 + a^2$ with $r \geq r_{\rm z}$ gives
\begin{align}
    G_\xi'(r) = 3 r^2 - 6 M r + r_{\rm z}^2 + 2a^2, \nonumber \\
    G_\xi''(r) = 6(r-M) > 0.
\end{align}
At $r_{\rm z}$, the slope of $G_\xi$ is also positive:
\begin{align}
    G_\xi'(r_{\rm z}) =2\bigl(2 r_{\rm z}^2 - 3Mr_{\rm z} + a^2\bigr)>0,
\end{align}
because $2 r^2 - 3 M r + a^2$ equals $r_+ (r_ + -M)>0$ at the horizon and increases throughout the exterior of the domain. Since the slope of $G_\xi$ continues to increase, $G_\xi$ remains positive above $r_{\rm z}$. Equation~\ref{eq:kerr-F-derivative} then gives $F_\xi'>0$ there. Together with $F_\xi'<0$ below $r_{\rm z}$, this establishes a single minimum at $r_{\rm z}$.

If $P_\xi$ has no exterior zero, it stays positive, so $F_\xi'$ has the same sign as $G_\xi$. A minimum can therefore occur only where $G_\xi$ crosses from negative to positive. At the horizon,
\begin{align}
    G_\xi(r_+) = -(r_+ - M) P_\xi(r_+) \leq 0,
\end{align}
while $G_\xi(r) \to +\infty$ as $r \to \infty$. Because $G_\xi'' = 6(r-M)>0$, its slope increases throughout the exterior. Starting from a strictly negative horizon value, $G_\xi$ can therefore cross zero only once, giving $F_\xi$ a single positive minimum. In the limiting case $P_\xi(r_+) = 0$, $G_\xi$ starts at zero with $G_\xi'(r_+)=2r_+(r_+-M)>0$. It then stays positive, and $F_\xi$ increases throughout the exterior.

The curve $F_\xi$ therefore meets any horizontal level at most twice in the exterior, and if two simple roots exist, the smaller lies on its decreasing part and the larger on its increasing part. At any exterior root, Equation~\ref{eq:kerr-radial-level-set} gives
\begin{align}
    \mathcal{R}'(R)=\Delta_R F_\xi'(R), \label{eq:kerr-slope-identity}
\end{align}
so a positive radial slope identifies a root on the increasing part of $F_\xi$. The curve continues to rise for every larger radius, so no larger root can block the inward ray. When the horizontal level reaches the minimum of $F_\xi$, the two roots merge, giving $\mathcal{R}'(R) = 0$ and $\mathcal{R}''(R) > 0$. Thus
\begin{align}
    R=T(\alpha,\beta) \quad\Longleftrightarrow\quad \mathcal{R}(R)=0, \qquad \mathcal{R}'(R)\geq0. \label{eq:kerr-branch-rule}
\end{align}
We can therefore construct each closest-approach contour by testing the slope on its candidate conic without comparing all radial roots at each image point. The points with $\mathcal{R}'(R) > 0$ form the physical arcs shown in Figure~\ref{fig:kerr-tiling}. At the boundary of the returning region, $\mathcal{R}'(R) = 0$, and the ray approaches an unstable spherical photon orbit asymptotically, giving a point on the critical curve.

\subsection{Constructing the boundary of the black hole shadow}
\label{sec:kerr-critical}

\begin{figure*}[t]
    \centering
    \includegraphics[width=0.9\linewidth]{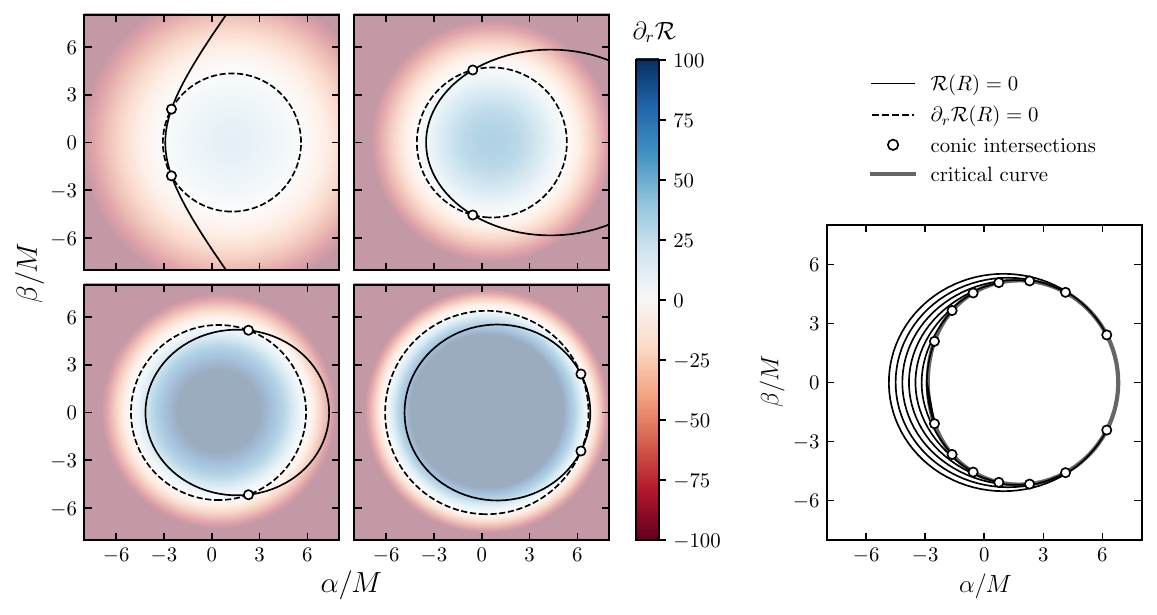}
    \caption{Construction of the Kerr critical curve from conic intersections for $a/M = 0.9$ and $\theta_o = 80^\circ$. Each left panel shows a candidate conic $\mathcal{R}(R) = 0$ as a solid curve and the derivative conic $\mathcal{R}'(R) = 0$ as a dashed curve. The background color gives $\mathcal{R}'(R)$, with blue positive and red negative. Along the solid arc inside the dashed curve, $\mathcal{R}'(R) > 0$, so inward rays first turn at $R$. This arc is the physical closest-approach contour. At its endpoints, white circles mark intersections satisfying $\mathcal{R}(R) = \mathcal{R}'(R) = 0$. The corresponding rays asymptotically approach unstable spherical photon orbits at $R$, where $\mathcal{R}''(R) > 0$. On the right, the physical conic arcs for several orbit radii are overlaid in black. Their endpoints trace the gray critical curve as $R$ varies.}
    \label{fig:kerr-conic-intersections}
\end{figure*}

To locate the critical curve, we find the image points where the outer radial root satisfies
\begin{align}
    \mathcal{R}(R)=0, \qquad \mathcal{R}'(R)=0, \label{eq:kerr-repeated-root}
\end{align}
where the derivative is taken at fixed constants of motion. These conditions select a root of multiplicity at least two and with exact multiplicity two when $\mathcal{R}''(R) \neq 0$ (Equation~\ref{eq:kerr-multiplicity-two}). On the image, the first condition defines the fixed-root conic $\mathcal{C}_0 = 0$. Differentiating Equation~\ref{eq:kerr-radial-expanded} and substituting the expressions for $\xi$ and $\eta$ in terms of $(\alpha, \beta)$ gives a second conic,
\begin{align}
    \mathcal{C}_1(\alpha,\beta;R) &=C_1(R)+L_1(R)\alpha \nonumber \\
    &\quad +A_1(R)\alpha^2-B_1(R)\beta^2=0, \label{eq:kerr-derivative-conic}
\end{align}
with
\begin{align}
    B_1(R) &= 2(R - M), \nonumber \\
    A_1(R) &= -2(R - M), \nonumber \\
    L_1(R) &= 4M a \sin\theta_o,
\end{align}
and
\begin{align}
    C_1(R) &= 4 R^3 + 2 R a^2(1+\cos^2\theta_o) +2 M a^2 \sin^2 \theta_o.
\end{align}
Since $A_1 = -B_1 < 0$ outside the horizon, the derivative conic is a circle with $\mathcal{R}'(R) > 0$ inside it. The panels of Figure~\ref{fig:kerr-conic-intersections} show how this circle selects physical closest-approach contours at several radii. In each panel, the portions of the solid fixed-root conic inside the dashed circle (the ${\cal C}_1$ conic) lie in the blue region and satisfy the local slope test.

At each radius $R$, the endpoints of the selected arcs lie at the intersections
\begin{align}
    \mathcal{C}_0(\alpha,\beta;R) =\mathcal{C}_1(\alpha,\beta; R)=0 \label{eq:kerr-two-conic-intersection}
\end{align}
of the two conics. As shown above, these exterior repeated roots have $\mathcal{R}''(R) > 0$, so they have exact multiplicity two. The photon's polar motion must also be allowed at the observer. Varying $R$ through the radii that satisfy this condition traces the critical curve.

As a consistency check, we recover the familiar Bardeen parametrization from these same conditions. Writing $P_R=P_\xi(R)$, the repeated-root equations become
\begin{align}
    P_R^2&=\Delta_R\kappa, & 4RP_R&=\Delta_R'\kappa,
\end{align}
At an exterior root, $P_R = 0$ would require $\kappa = 0$ because $\Delta_R > 0$. Equation~\ref{eq:kerr-kappa-screen} would then give $\beta = 0$, $\alpha = -a\sin\theta_o$, and $\xi = a\sin^2\theta_o$, implying $P_R = R^2+a^2\cos^2\theta_o>0$. Thus $P_R \neq 0$ for a ray reaching the observer. Eliminating $\kappa$ between the two equations and dividing by $P_R$ gives
\begin{align}
    P_R &= \frac{2 R \Delta_R}{R - M}, & \kappa_c(R)&=\frac{4 R^2 \Delta_R}{(R - M)^2},
\end{align}
where the subscript $c$ denotes values for rays on the critical curve, which approach unstable spherical photon orbits asymptotically. For $a \neq 0$, using $P_R = R^2+a^2-a\xi$ and $\eta = \kappa - (\xi - a)^2$ gives
\begin{align}
    \xi_c(R) &= \frac{R^2(R-3M)+a^2(R+M)}{a(M-R)}, \label{eq:kerr-xi-critical} \\
    \eta_c(R) &= \frac{R^3\left[4Ma^2-R(R-3M)^2\right]} {a^2(R-M)^2}. \label{eq:kerr-eta-critical}
\end{align}
Substituting these constants into the relations for $\alpha$ and $\beta$ gives the critical curve
\begin{align}
    \alpha_c(R)&=-\frac{\xi_c(R)}{\sin\theta_o}, \nonumber \\
    \beta_c(R)&=\pm\sqrt{ \eta_c(R)+a^2\cos^2\theta_o -\xi_c(R)^2\cot^2\theta_o}. \label{eq:kerr-critical-parametrization}
\end{align}
The observer's inclination sets the allowed range of $R$: the polar motion must be allowed at $\theta_o$, which requires $\beta_c$ to be real. At each allowed radius, the two signs of $\beta_c$ give the upper and lower endpoints of the selected closest-approach contour, where the radial root has exact multiplicity two. For positive spin, the smallest visible radii contribute to the prograde, negative-$\alpha$ side of the critical curve, while progressively larger radii extend toward the retrograde side. The orbit radius $R$ therefore records how deeply the corresponding critical rays approach the black hole. As $a \to 0$, the visible spherical-orbit interval collapses to $R = 3M$, so $R$ no longer distinguishes points around the critical curve. At $a=0$, the two conics share the critical circle $\alpha^2+\beta^2 = 27 M^2$.

\subsection{Images for observers at finite radius}
\label{sec:kerr-finite-observers}

We now consider the image seen by an observer closer to the black hole. As before, each ray arrives from a direction on the observer's local sky, which we project onto a flat image to draw the closest-approach contours. We use a zero angular momentum observer (ZAMO) outside the horizon and away from the rotation axis \cite{bardeen_1972_kerr,grenzebach_2014_kerrnewmannutshadows}. The four-velocity of the observer is proportional to
\begin{align}
    \partial_t+\Omega_{\rm Z}\partial_\phi, \qquad \Omega_{\rm Z}=-\frac{g_{t\phi}}{g_{\phi\phi}},
\end{align}
and satisfies $u_\phi = 0$ by definition. For photons with nonzero Killing energy $E = -k_t$, we define finite-radius impact coordinates $(\widehat\alpha, \widehat\beta)$ by dividing the azimuthal and polar components of the photon momentum in the observer's local orthonormal frame by $E$. The constants of motion then satisfy
\begin{align}
    \xi &= p_0+p_1\widehat\alpha, \nonumber \\
    \eta &= q_0+q_1\widehat\alpha +q_2\widehat\alpha^2+q_3\widehat\beta^2, \label{eq:kerr-finite-impact-map}
\end{align}
where the coefficients $p_i$ and $q_i$ are fixed by the observer event and Kerr parameters. Importantly, these coefficients are the same for every image point and trial radius. Since the radial potential is quadratic in $\xi$ and linear in $\eta$, both $\mathcal{R}(R)=0$ and $\mathcal{R}'(R)=0$ remain conics in these impact coordinates.

Dividing the azimuthal and polar momentum components by the locally measured photon energy gives the horizontal and vertical coordinates of an orthographic image. This projection maps a hemisphere of the local sky onto a disk along parallel lines. For a distant observer, these coordinates become small angular offsets. With the axes oriented as in the distant image, scaling these offsets by the observer distance recovers Bardeen's $(\alpha, \beta)$. The energy measured by the ZAMO is
\begin{align}
    \omega_{\rm Z} = -k_\mu u_{\rm Z}^\mu =E \,u_{\rm Z}^t(1-\Omega_{\rm Z}\xi).
\end{align}
Since $\xi = p_0 + p_1\widehat\alpha$, the energy ratio has the form $\omega_{\rm Z}/E = s_0 + s_1\widehat\alpha$. Dividing both impact coordinates by this ratio gives the transverse direction cosines $(X, Y)$ that serve as orthographic image coordinates:
\begin{align}
    X &= \frac{\widehat\alpha}{s_0+s_1\widehat\alpha},
    &
    Y &= \frac{\widehat\beta}{s_0+s_1\widehat\alpha}.
    \label{eq:kerr-projective-camera}
\end{align}
where $s_0$ and $s_1$ are fixed by the observer event and Kerr parameters. For $E \neq 0$, this projective change of coordinates preserves both the fixed-root and derivative conics.\footnote{A pinhole camera places each image point where its viewing ray intersects a plane, and this can change the contours' conic form. In the distant Kerr limit, however, both projections reduce to small angular offsets and recover Bardeen impact coordinates after scaling by observer distance.}

For a returning ray traced inward from an observer at $r_o$, the closest approach lies between the observer and the horizon, $r_+ < R < r_o$. The radial slope test then selects the physical arcs of each candidate conic. Inside the ergoregion, directions with $E = 0$ require separate coordinates because the impact coordinates divide by $E$. We show in Appendix~\ref{app:kerr-extensions} that changing the observer's motion at the same event changes the apparent ray directions while preserving each ray's turning radius.

\subsection{From Kerr to the PD family}

In Kerr, we have shown that the closest-approach contours and the black hole shadow boundary follow from the same family of conics. Depending on the trial radius and viewing geometry, the candidate conics are hyperbolas, parabolas, or ellipses. Equation~\ref{eq:kerr-branch-rule} can then be used to select the physical closest-approach arcs from the full candidate conic curves. The intersections with the derivative conic give image points whose rays approach unstable spherical photon orbits, provided their polar motion is allowed. As $R$ varies, these points trace the critical curve. The impact coordinates and orthographic projection above preserve the conics at finite observer radius, with the closest-approach map restricted to initially inward rays.

In Kerr, the single minimum or monotonic rise of $F_\xi$ fixes the order of the exterior roots and explains why the local slope test is sufficient. The next section establishes when this same local slope test identifies the first turn of an inward ray in the broader PD family of spacetimes by using a restriction on the radial potential imposed by the field equations. Appendix~\ref{app:kerr-extensions} treats extremal Kerr, the Schwarzschild limit, and changes in the image caused by observer motion.

\section{Pleba\'{n}ski--Demia\'{n}ski spacetimes}
\label{sec:pd}

\begin{table*}[t]
    \centering
    \caption{Radial root selection in PD, summarizing the results and their assumptions from Sections~\ref{sec:pd-positive-radius}--\ref{sec:mixed-sign}.}
    \label{tab:pd-radial-results}
    \small
    \begin{tabular*}{\textwidth}{@{}p{0.45\textwidth}@{\extracolsep{\fill}}p{0.51\textwidth}@{}}
        \hline\hline
        Radial domain & Conclusion \\
        \hline
        $m>0$; a connected positive-radius interval with $\Delta_r>0$ and radial motion allowed at its endpoints
        & The local slope test selects the first root reached by an inward ray without assuming a single minimum of the radial root curve. \\
        $m>0$, $\ell=0$; an allowed outer radial region
        & The radial root curve is monotonic or has one interior minimum, so each ray has at most two distinct radial roots there. \\
        $m>0$; a connected interval with $\Delta_r>0$ crossing $r=0$ and with radial motion allowed at its boundaries
        & The local test can retain a blocked root. Each failure requires one negative root. Any blocked positive repeated root has $\mathcal{R}''<0$, so it cannot be approached from neighboring radii. \\
        \hline\hline
    \end{tabular*}
\end{table*}

We extend our analysis beyond Kerr to the PD family of metrics in order to ask whether the local slope test survives when charge, acceleration, cosmological curvature, and NUT charge change the radial motion. These extensions retain the constants of motion used in Kerr: stationarity and axisymmetry give the photon energy $E=-k_t$ and axial angular momentum $L_z=k_\phi$, while a hidden symmetry supplies the separation constant $K$. We can therefore continue to treat the photon's radial and angular motion separately.

The PD family extends Kerr within the exact solutions of the Einstein--Maxwell equations with a cosmological constant, which constrain how the additional parameters can change the geometry. In PD, tidal gravity and the electromagnetic field select the same pair of principal null directions, as described above. With this alignment, the field equations restrict the radial metric function to a polynomial of degree at most four, with coefficients tied to the angular geometry and physical parameters of the spacetime \cite{plebanski_1976_pdfamily}. These restrictions will allow us to select the physical closest-approach contours from the candidate conics.

We write the PD metric in the Boyer--Lindquist coordinates $(t,r,\vartheta,\phi)$ and physical parametrization used by Grenzebach, Perlick, and L\"{a}mmerzahl (GPL) \cite{grenzebach_2015_gplpd}, which provides both the separated photon equations and the local observer frame used to construct the image. In the GPL form, the line element is
\begin{align}
    ds^2 &= \frac{1}{\Omega^2}\biggl[
    -\frac{\Delta_r}{\Sigma}(dt-\chi\,d\phi)^2
    \nonumber \\
    &\qquad +\frac{\Sigma}{\Delta_r}\,dr^2
    +\frac{\Sigma}{\Delta_\vartheta}\,d\vartheta^2 \nonumber \\
    &\qquad +\frac{\Delta_\vartheta\sin^2\vartheta}{\Sigma}
    \bigl(a\,dt-(\Sigma+a\chi)\,d\phi\bigr)^2\biggr],
    \label{eq:pd-metric}
\end{align}
where
\begin{align}
    \Sigma &= r^2+(\ell+a\cos\vartheta)^2, \nonumber \\
    \chi &= a\sin^2\vartheta-2\ell(\cos\vartheta+C), \nonumber \\
    \Omega &= 1-\frac{\alpha_{\rm acc}}{\omega}(\ell+a\cos\vartheta)r,
    \qquad \omega=\sqrt{a^2+\ell^2}, \label{eq:pd-metric-functions} \\
    \Delta_\vartheta &= 1-a_3\cos\vartheta-a_4\cos^2\vartheta, \nonumber \\
    \Delta_r &= b_0+b_1r+b_2r^2+b_3r^3+b_4r^4.
    \label{eq:pd-metric-polynomials}
\end{align}
The parameters $m$ and $a$ describe mass and rotation and reduce to $M$ and $J/M$, respectively, in Kerr. The additional parameters introduced by the PD metric are a NUT charge $\ell$, acceleration $\alpha_{\rm acc}$, cosmological constant $\Lambda$, and charge combination $\beta_{\rm ch}= q_e^2 + q_m^2$, where $q_e$ and $q_m$ are the electric and magnetic charges. The parameter $C$ specifies the placement of the NUT string singularity on the rotation axis.\footnote{This parameter enters the metric only when $\ell \neq 0$. The choice $C = 0$ distributes the NUT string singularity symmetrically between the two parts of the axis, while $C = \pm 1$ places it on one side.}

The additional parameters change the domains in which photons are allowed to propagate. The acceleration parameter describes acceleration along the symmetry axis; it changes both the angular geometry and the conformal boundary $\Omega=0$, and it can introduce an acceleration horizon. In accelerating black hole solutions, small circles around the two sides of the symmetry axis can generically have different ratios of circumference to proper radius. Choosing the azimuthal period on one side to make this ratio $2\pi$ leaves a so-called conical defect on the other, which is often interpreted as a cosmic string pulling the black hole or a strut pushing it. The NUT parameter couples the time and azimuthal directions, giving the spacetime an axial twist even when ordinary rotation vanishes. Charge and the cosmological constant also change $\Delta_r$, whose roots locate radial horizons. The same function enters the photon radial potential, so these parameters also affect where rays can turn.

The same spacetime parameters determine the coefficients of both $\Delta_r$ and $\Delta_\vartheta$. To express these coefficients in a more compact and legible form, we first define the auxiliary combinations
\begin{align}
    k &= (a^2-\ell^2)
    \frac{1+2\frac{\alpha_{\rm acc}}{\omega}\ell m
    -3\ell^2\bigl(\frac{\alpha_{\rm acc}^2}{\omega^2}\beta_{\rm ch}
    +\frac{\Lambda}{3}\bigr)}
    {1+3\frac{\alpha_{\rm acc}^2}{\omega^2}\ell^2(a^2-\ell^2)}, \nonumber \\
    \zeta &= \frac{\alpha_{\rm acc}^2}{\omega^2}(k+\beta_{\rm ch})
    +\frac{\Lambda}{3}.
    \label{eq:pd-coefficient-combinations}
\end{align}
\begin{align}
    a_3 &= 2\frac{\alpha_{\rm acc}}{\omega}am-4a\ell\zeta,
    \qquad a_4 = -a^2\zeta, \nonumber \\
    b_0 &= k+\beta_{\rm ch}, \qquad b_1 = -2m, \nonumber \\
    b_2 &= \frac{k}{a^2-\ell^2}
    +4\frac{\alpha_{\rm acc}}{\omega}\ell m
    -(a^2+3\ell^2)\zeta, \nonumber \\
    b_3 &= -2\frac{\alpha_{\rm acc}}{\omega}\biggl[
    \frac{k\ell}{a^2-\ell^2}
    -(a^2-\ell^2)\biggl(\frac{\alpha_{\rm acc}}{\omega}m-\ell\zeta\biggr)
    \biggr], \nonumber \\
    b_4 &= -\frac{\alpha_{\rm acc}^2}{\omega^2}k-\frac{\Lambda}{3}.
    \label{eq:pd-metric-coefficients}
\end{align}
The ratios $k/(a^2-\ell^2)$ are evaluated after canceling the common factor in Equation~\ref{eq:pd-coefficient-combinations}. The simultaneous $a = \ell=0$ limit is obtained from the $\ell=0$ expressions as $a \to 0^+$. Setting $\ell = \alpha_{\rm acc} = \beta_{\rm ch} = \Lambda = 0$ recovers the Kerr metric, with $\Omega = \Delta_\vartheta = 1$ and $\Delta_r = r^2 - 2mr + a^2$. Across the PD family, the coefficient $b_1 = -2m$ remains fixed by the mass. This is the restriction that will control the ordering of the photon radial roots.

First, we will show that the constants of motion still give fixed-root conics in adapted image coordinates. We will then show that, for $m > 0$, the restriction $b_1 = -2m$ makes the local slope test sufficient to select the physical radial turn on the positive-radius intervals specified below. Table~\ref{tab:pd-radial-results} summarizes this result and its scope. In the limit that NUT charge vanishes, we also recover the Kerr-like behavior of $F_\xi(r)$ as a function of radius, showing that it is monotonic or has a single interior minimum. The final two subsections examine how an inner candidate root can be hidden by a larger turn when the radial interval crosses zero or the geometric alignment is relaxed. In the latter case, even an unstable photon orbit at positive radius can be hidden.

\subsection{Conics on a finite observer's image}
\label{sec:pd-screen}

In PD, a cosmological or acceleration horizon can bound the observer's stationary region, forcing any observer to be positioned at finite radius, so we therefore consider the image formed from photon directions measured locally. GPL related these directions to the photon constants of motion to construct the shadow boundary \cite{grenzebach_2015_gplpd}. We use their local frame to construct the candidate closest-approach contours, treating rays with finite turning points alongside critical rays that approach unstable photon orbits. We restrict this PD image construction to photons with positive Killing energy $E=-k_t>0$ and define the normalized constants
\begin{align}
    \xi=\frac{L_z}{E}, \qquad \kappa=\frac{K}{E^2},
\end{align}
with the same conventions as Equation~\ref{eq:kerr-constants} in the Kerr limit. The radial potential can then be written as
\begin{align}
    \mathcal{R}(r) = \left[A_r(r)-a\xi\right]^2-\kappa\Delta_r(r), \label{eq:pd-radial}
\end{align}
where the angular terms in $\Sigma+a\chi$ cancel, giving
\begin{align}
    A_r(r) = \Sigma + a\chi = r^2 + a^2 + \ell^2 - 2a \ell C.
\end{align}
The corresponding angular potential is
\begin{align}
    \Theta(\vartheta) =\Delta_\vartheta\kappa -\frac{(\chi-\xi)^2}{\sin^2\vartheta}.
\end{align}
As in Kerr, the ray can propagate only where $\mathcal{R} \geq 0$ and $\Theta \geq 0$.

To express these potentials in image coordinates, we consider an observer at $(r_o, \vartheta_o)$, away from the rotation axis, where
\begin{align}
    \Delta_{r,o}&>0, & \Delta_{\vartheta o}&>0, & \Sigma_o&>0, \nonumber \\
    \Omega_o&>0, & \sin\vartheta_o&\neq0.
\end{align}
GPL's orthonormal local frame (tetrad), evaluated at the observer's position $(r_o, \vartheta_o)$, is
\begin{align}
    \boldsymbol{e}_{(0)} &= \frac{\Omega_o}{\sqrt{\Sigma_o\Delta_{r,o}}}
    \bigl[A_r(r_o)\partial_t+a\partial_\phi\bigr], \nonumber \\
    \boldsymbol{e}_{(1)} &= \Omega_o\sqrt{\frac{\Delta_{\vartheta o}}{\Sigma_o}}\,\partial_\vartheta, \nonumber \\
    \boldsymbol{e}_{(2)} &= -\frac{\Omega_o}{\sqrt{\Sigma_o\Delta_{\vartheta o}}\sin\vartheta_o}
    \bigl[\partial_\phi+\chi_o\partial_t\bigr], \nonumber \\
    \boldsymbol{e}_{(3)} &= -\Omega_o\sqrt{\frac{\Delta_{r,o}}{\Sigma_o}}\,\partial_r.
    \label{eq:pd-observer-tetrad}
\end{align}
Here, the observer has four-velocity $\boldsymbol{e}_{(0)}$ and measures photon momentum along the three spatial axes. The radial axis $\boldsymbol{e}_{(3)}$ is chosen to point inward, while $\boldsymbol{e}_{(1)}$ and $\boldsymbol{e}_{(2)}$ give the transverse polar and azimuthal directions. This tetrad is adapted to the principal null directions and reduces in Kerr to the Carter frame, rather than the ZAMO frame used in Section~\ref{sec:kerr-finite-observers}, and its observer has angular velocity $d\phi/dt = a/A_r(r_o)$. We choose this frame because it simplifies the image-coordinate relations below. For other observer velocities, the same conic representation can be recovered by correcting the measured photon directions for aberration into this reference frame before applying the image projection, as described in Appendix~\ref{app:abbreation}.

We divide the momentum components along these transverse axes by the Killing energy to define the impact coordinates:
\begin{align}
    X &= -\frac{k_\mu e_{(2)}^\mu}{E}, & Y&=-\frac{k_\mu e_{(1)}^\mu}{E}.
\end{align}
The minus signs orient the image toward the direction from which the photon arrives. The constants of motion then satisfy
\begin{align}
    \xi&=\chi_o+p_1X, & \kappa&=q_o(X^2+Y^2), \label{eq:pd-screen-map}
\end{align}
with
\begin{align}
    p_1&= \frac{\sqrt{\Sigma_o\Delta_{\vartheta o}}\sin\vartheta_o} {\Omega_o}, & q_o&=\frac{\Sigma_o}{\Omega_o^2}.
\end{align}
These coefficients are fixed by the spacetime and observer and are the same for every image point and trial radius. Since $\xi$ is linear in $X$ up to a constant offset and $\kappa$ is quadratic in $(X, Y)$, substituting them into $\mathcal{R}(R)=0$ gives the candidate conic $\mathcal{C}_0=0$:
\begin{align}
    \left[A_R - a(\chi_o+p_1X)\right]^2 - \Delta_Rq_o(X^2+Y^2) = 0, \label{eq:pd-root-conic}
\end{align}
where $A_R = A_r(R)$ and $\Delta_R = \Delta_r(R)$. Each conic identifies the image points for which $R$ is a candidate turning radius.

To place these curves on the observer's local sky, we instead normalize the transverse momentum by the locally measured photon energy, $E_{\rm loc} = -k_\mu e_{(0)}^\mu$. The ratio between energies is
\begin{align}
    W(X) =\frac{E_{\rm loc}}{E} = \frac{\Omega_o}{\sqrt{\Sigma_o\Delta_{r,o}}} \left(\Sigma_o-ap_1X\right),
\end{align}
and the orthographic image coordinates are therefore $x = X/W(X)$ and $y = Y/W(X)$. Since $W(X)$ is linear in $X$ up to a constant offset, this change of coordinates is projective and preserves conics. As in the finite-radius Kerr image, the same transformation applies at every trial radius and preserves the entire family of candidate conics. For positive locally measured energy and nonzero radial momentum, we require
\begin{align}
    W>0, \qquad X^2+Y^2<W^2, \label{eq:pd-inward-screen}
\end{align}
where $W > 0$ ensures $E_{\rm loc} > 0$ for the $E > 0$ light rays we consider. The strict inequality places the image point inside the unit disk $x^2 + y^2 < 1$ and gives nonzero radial momentum. We choose its sign so that tracing backward initially moves inward. The arriving photon then has ${\rm d}r/{\rm d}s > 0$, with affine parameter $s$ increasing toward the future. Following the ray toward decreasing $s$ takes it toward smaller $r$.

The image conditions also ensure that radial and angular motion are allowed at the observer, since
\begin{align}
    W^2-X^2-Y^2 &= \frac{\Omega_o^2}{\Sigma_o\Delta_{r,o}} \mathcal{R}(r_o), \nonumber \\
    \Theta(\vartheta_o) &= \frac{\Sigma_o\Delta_{\vartheta o}}{\Omega_o^2}Y^2.
\end{align}
Thus $\mathcal{R}(r_o) > 0$, $\kappa \geq 0$, and the local polar momentum is real. To obtain a physical closest-approach contour, however, we must still select the image points whose first accessible radial root is $R$, the ray's angular motion must also remain allowed, and its path must stay regular within the chosen $\Omega>0$ region. We next show when the local radial slope test is sufficient to identify this first root.

\subsection{Selecting the correct radial root}
\label{sec:pd-positive-radius}

In Kerr, the shape of the radial root curve made the local slope test sufficient to identify the first radial turn.\footnote{For Kerr, \citet{paganini_2016_nullgeodesics} use the signs of the radial polynomial's coefficients and its behavior at the horizon and infinity to constrain the number of exterior roots.} In PD, the same local selection rule follows from a restriction on the radial potential imposed by the field equations, under the positive-radius and endpoint conditions specified below. For fixed $\xi$, the radial potential takes the form
\begin{align}
    \mathcal{R}(r) &= (r^2+c_\xi)^2 - \kappa\Delta_r(r), \nonumber \\
    c_\xi &= a^2 + \ell^2 - 2a\ell C - a\xi.
\end{align}
The squared term contains no term linear in $r$, so the linear coefficient of the photon potential comes entirely from $\Delta_r$. The field equations fix this coefficient through the mass. Denoting the coefficient of $r$ in a polynomial $P(r)$ by the expression $[r]P$, we have
\begin{align}
    [r]\Delta_r = -2m, \qquad [r]\mathcal{R} = 2m\kappa. \label{eq:pd-linear-coefficient}
\end{align}

For $m > 0$, the local slope test selects the first radial root on positive-radius intervals where $\Delta_r > 0$ and radial motion is allowed at both endpoints. For a finite observer at $r_o$, the inner endpoint lies at $0 < r_{\rm in} < r_o$, and the ray has nonzero radial momentum at the observer:
\begin{align}
    \mathcal{R}(r_{\rm in})\geq0, \qquad \mathcal{R}(r_o)>0.
\end{align}
When the inner endpoint is a horizon, $\Delta_r(r_{\rm in}) = 0$, the inner radial condition holds automatically because
\begin{align}
    \mathcal{R}(r_{\rm in}) = (r_{\rm in}^2+c_\xi)^2\geq0.
\end{align}
If the potential has a root between $r_{\rm in}$ and the observer, the inward ray first approaches the outermost root, whose radius is
\begin{align}
    T = \max\{r\in(r_{\rm in},r_o):\mathcal{R}(r)=0\}.
\end{align}
The potential's slope at a candidate root is enough to determine whether that root sets the ray's closest approach. For any trial radius $R \in (r_{\rm in}, r_o)$,
\begin{align}
    R=T \quad\Longleftrightarrow\quad \mathcal{R}(R)=0, \qquad \mathcal{R}'(R)\geq0. \label{eq:pd-positive-branch}
\end{align}
For $\mathcal{R}'(T) > 0$, the ray reaches $r = T$ and turns outward because radial motion is forbidden immediately inside this radius. For $\mathcal{R}'(T) = 0$, the root has exact multiplicity two and
\begin{align}
    \mathcal{R}''(T)>0,
\end{align}
so the ray asymptotically approaches an unstable photon orbit at $r = T$, provided its path satisfies the regularity conditions in Appendix~\ref{app:pd-branch-proof}.

For $\kappa > 0$, the local test could fail only if the inward ray approached an outer root before an inner root that also passes the test. If both roots are simple, radial motion is allowed just outside each one and forbidden just inside. The forbidden region inside the outer root must end before the inner root, so another root lies between them. A fourth root is needed below the inner root, where radial motion must become allowed again by the inner endpoint. Repeated roots lead to the same minimum count of four positive roots when multiplicity is included.

Since $\mathcal{R}$ has degree at most four, these are all of its roots. The potential is positive at the observer, outside all four roots, so its leading coefficient $A$ must also be positive. Four positive roots would then require a negative coefficient of $r$. For roots $r_i > 0$, Vieta's relation gives
\begin{align}
    [r]\mathcal{R} =-A\sum_{i<j<k}r_ir_jr_k<0, \qquad A>0, \label{eq:pd-vieta-contradiction}
\end{align}
contradicting the positive coefficient $2m\kappa$ in Equation~\ref{eq:pd-linear-coefficient}. For $\kappa=0$, the potential is $\mathcal{R} = (r^2+c_\xi)^2$, whose positive root, if present, is unique. Appendix~\ref{app:pd-branch-proof} gives the full root count, including repeated roots.

The same result applies when the positive-radius interval extends to infinity and radial motion remains allowed sufficiently far out. In this case, we can apply the finite-interval result at an outer endpoint beyond all finite roots. On an interval
\begin{align}
    I = (r_{\rm in},\infty),
\end{align}
with $r_{\rm in} > 0$, $\Delta_r > 0$ throughout $I$, and $\mathcal{R}(r_{\rm in}) \geq 0$, we require
\begin{align}
    \mathcal{R}(r)>0 \qquad \text{for all sufficiently large }r.
\end{align}
If the interval contains a finite root, its largest root $T_\infty$ is therefore selected by the same local test: for $R \in I$,
\begin{align}
    R = T_\infty \quad\Longleftrightarrow\quad \mathcal{R}(R)=0, \qquad \mathcal{R}'(R)\geq0. \label{eq:pd-unbounded-branch}
\end{align}

The remaining question of whether radial motion remains allowed at large radius is answered by examining the highest powers of $r$ in the radial potential. Using the coefficients of $\Delta_r$ defined in Equation~\ref{eq:pd-metric-polynomials}, we have
\begin{align}
    \mathcal{R}(r) &= (1-\kappa b_4)r^4-\kappa b_3r^3 \nonumber \\
    &\quad +(2c_\xi-\kappa b_2)r^2+2m\kappa r +(c_\xi^2-\kappa b_0). \label{eq:pd-tail}
\end{align}
The coefficient of the highest nonzero power of $r$ determines the sign at large radius. When $b_4 = 0$, the positive $r^4$ term guarantees that radial motion is allowed sufficiently far out. For $b_4>0$, radial motion remains allowed there when
\begin{align}
    \kappa < \frac{1}{b_4},
\end{align}
and is forbidden for $\kappa > 1/b_4$. At $\kappa = 1/b_4$, the quartic term vanishes and the highest remaining nonzero term determines whether radial motion is allowed. For $b_4 < 0$, the photon potential is positive at large radius for $\kappa \geq 0$. The metric function $\Delta_r$ is then negative there, so this case falls outside the unbounded $\Delta_r > 0$ intervals covered by the result above.

These large-radius conditions describe radial propagation, but to obtain a distant image, we must also specify how the observer approaches the spacetime boundary and how the image coordinates are scaled. In accelerating solutions, the conformal boundary $\Omega = 0$ generally depends on angle, so the limit requires a choice of boundary location and observer frame. The observer and ray must remain within the chosen $\Omega > 0$ region. Appendix~\ref{app:boundary-screens} describes the standard asymptotically flat and nonaccelerating AdS limits and the additional choices required with acceleration.

The local slope test $\mathcal{C}_1 \geq 0$ therefore identifies the points on each candidate conic $\mathcal{C}_0 = 0$ whose rays have $R$ as their first radial root. The physical arcs correspond to rays traced initially inward that can approach $R$ while their polar motion remains allowed and their path stays regular within the chosen $\Omega > 0$ region. Varying $R$ then builds the returning image directly from the root and derivative conics, without comparing the radial roots at every image point.

\subsection{Radial root structure without NUT charge}
\label{sec:no-nut-one-well}

Section~\ref{sec:pd-positive-radius} identifies the physical turning radius without assuming a particular shape for the radial root curve. When NUT charge vanishes, we can also determine this shape. For $m > 0$ and $\ell = 0$, the radial root curves in the outer regions we consider are monotonic or have a single minimum, as in Kerr. Each ray then has at most two distinct radial roots there, with the outer root giving the closest approach of an inward returning ray. In this limit, the radial coefficients satisfy
\begin{align}
    b_0 &= a^2 + \beta_{\rm ch} \geq 0, \nonumber \\
    b_1 &= -2m < 0, \nonumber \\
    b_3 &= 2m\alpha_{\rm acc}^2 \geq 0, \nonumber \\
    b_4 &= - \left( \alpha_{\rm acc}^2 + \frac{\Lambda}{3} \right).
    \label{eq:no-nut-coefficients}
\end{align}
These outer regions begin at a positive-radius horizon and have $\Delta_r > 0$ throughout. Depending on the sign of $q \equiv \alpha_{\rm acc}^2 + \Lambda/3$, they either end at another horizon or extend to arbitrarily large radius. For $q > 0$, the finite interval $I = (r_{\rm in}, r_{\rm out})$ lies between the two outermost distinct positive horizon radii, with $0 < r_{\rm in} < r_{\rm out}$ and $\Delta_r(r_{\rm in}) = \Delta_r(r_{\rm out}) = 0$. For $q \leq 0$, the interval $I = (r_{\rm in}, \infty)$ lies beyond the largest positive horizon radius, where $\Delta_r(r_{\rm in}) = 0$. Coincident horizons are included whenever the selected interval retains nonzero width. This leaves four possibilities:
\begin{enumerate}
    \item a finite interval between the horizons when $\alpha_{\rm acc}^2 + \Lambda/3 > 0$;
    \item an asymptotically flat exterior when $\alpha_{\rm acc} = 0 = \Lambda$, with a quadratic $\Delta_r$;
    \item an unbounded radial interval when $\alpha_{\rm acc} \neq 0$ and $\Lambda = -3\alpha_{\rm acc}^2$, so the $r^4$ term cancels and $\Delta_r$ has leading term $2m \alpha_{\rm acc}^2 r^3$;
    \item an unbounded radial interval with a quartic $\Delta_r$ when $\alpha_{\rm acc}^2 + \Lambda/3 < 0$.
\end{enumerate}
In each of these outer intervals, the radial root curve
\begin{align}
    F_\xi(r) = \frac{(r^2+a^2-a\xi)^2}{\Delta_r(r)}
\end{align}
has at most one interior minimum and no interior maximum. The radial roots occur at the intersections $F_\xi(r) = \kappa$, and a horizontal line can intersect the curve at most twice. For rays that return or whose radial motion remains allowed at arbitrarily large radius, a root on the increasing side of the curve is unique and is the first approached by an inward ray from larger radius. The curve's shape thus independently recovers the local slope test.

Appendix~\ref{app:no-nut} proves this result, including cases where the minimum lies at a zero of the numerator, horizons coincide, or the numerator and denominator share a root. Applying this result to accelerating black holes also requires the global geometric conditions discussed in Section~\ref{sec:discussion}.

\subsection{Why the positive-radius restriction matters}
\label{sec:mixed-sign}

The results described thus far have considered radial regions confined to $r > 0$. However, in general, the radial coordinate $r$ does not necessarily measure proper distance, and $r = 0$ need not be a singular surface. Indeed, some PD geometries extend through regular points at $r = 0$ to negative values of $r$ \cite{grenzebach_2015_gplpd}: in a nonrotating spacetime with NUT charge, the metric factor $\Sigma = r^2 + \ell^2$ remains finite and positive as $r$ passes through zero. If the connected region with $\Delta_r > 0$ extends across zero, we now show that a negative radial root can allow an inner positive root to pass the local slope test even though the ray turns at a larger radius.

\begin{figure*}[t]
    \centering
    \includegraphics[width=\linewidth]{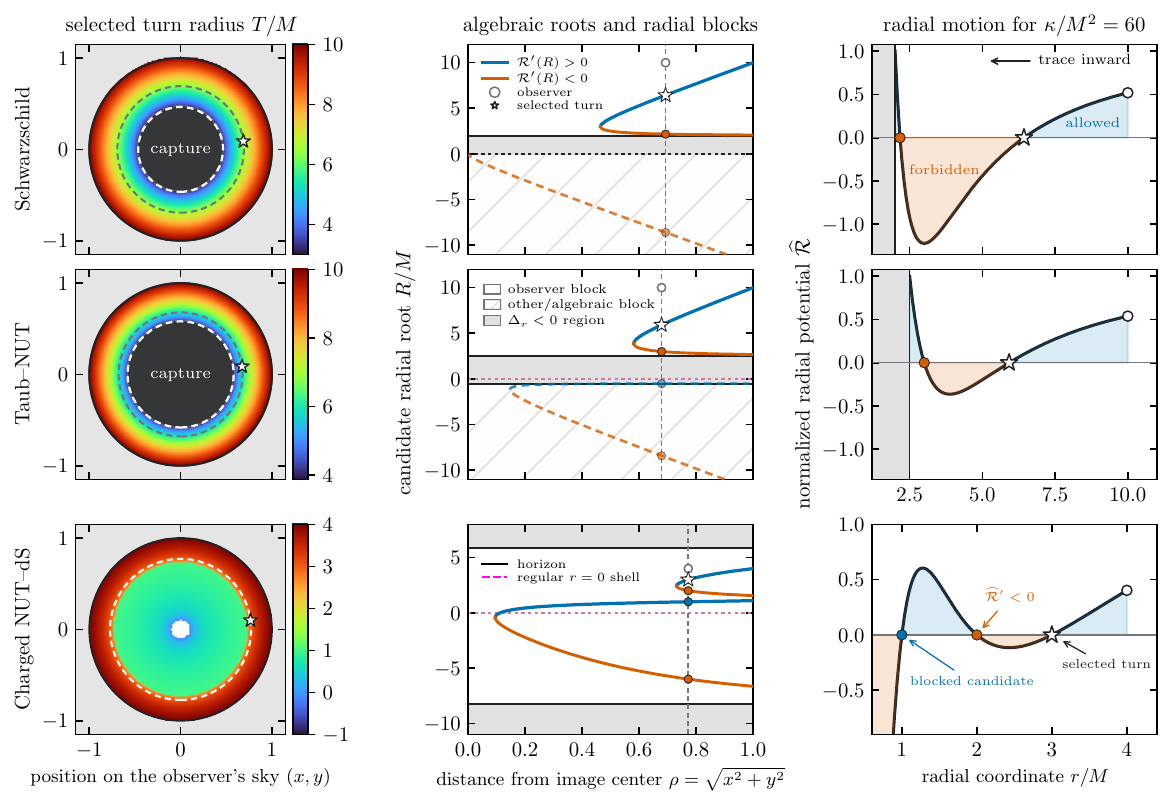}
    \caption{Closest approach for inward rays in Schwarzschild (top), pure Taub--NUT (middle), and the charged NUT--de Sitter example in Equation~\ref{eq:nut-example} (bottom). The left panels show closest-approach maps, and the center panels follow the radial roots as the distance from the image center varies. Blue and orange root curves have positive and negative $\mathcal{R}'$, respectively. The right panels show the normalized radial potential $\widehat{\mathcal{R}} = \mathcal{R}/(r^2 + \ell^2)^2$ for the selected ray with $\kappa/M^2 = 60$. This normalization preserves the roots and the sign of the potential over the plotted ranges. The white center in the bottom-left panel contains rays that leave the connected $\Delta_r > 0$ region containing the observer without turning. In Schwarzschild, the singularity at $r = 0$ prevents rays from reaching negative radius. In pure Taub--NUT, an intervening $\Delta_r < 0$ region separates the negative-radius roots from the observer. For both selected rays, the only positive-slope root in the observer's connected $\Delta_r > 0$ region gives the closest approach. In the charged example, however, the observer and all four roots lie in the same connected $\Delta_r > 0$ region. Both $R = M$ and $R = 3M$ satisfy $\mathcal{R}'(R) > 0$. An inward ray from $r_o = 4M$ reaches $R = 3M$ first, and the forbidden interval $2M < r < 3M$ prevents it from reaching $R = M$. The slope test therefore accepts the simple root at $R = M$ even though the ray turns at $R = 3M$.}
    \label{fig:nut-counterexample}
\end{figure*}

Figure~\ref{fig:nut-counterexample} compares the charged NUT--de Sitter example constructed below with Schwarzschild and pure Taub--NUT (the nonrotating, nonaccelerating vacuum member of PD with nonzero NUT charge and $\Lambda = 0$). In pure Taub--NUT, the ordinary exterior remains confined to positive radius, so NUT charge alone does not cause the local test to fail.

In the charged example, however, the horizons enclose a single region with $\Delta_r > 0$ that extends through $r = 0$. To understand how this changes the possible radial roots, we consider positive mass, $m > 0$, and two horizons on opposite sides of zero:
\begin{align}
    & r_{\rm in} < 0 < r_{\rm out}, \nonumber \\ 
    & \Delta_r(r_{\rm in}) = \Delta_r(r_{\rm out}) = 0, \nonumber \\
    & \Delta_r > 0\quad\text{on }(r_{\rm in},r_{\rm out}),
\end{align}
with radial motion allowed at both horizons,
\begin{align}
    \mathcal{R}(r_{\rm in}) \geq 0, \qquad \mathcal{R}(r_{\rm out}) > 0.
\end{align}
An inward ray from an observer at $0 < r_o < r_{\rm out}$ starts in an allowed region, with $\mathcal{R}(r_o) > 0$, and its closest approach is set by the first radial root it encounters.

Suppose a positive root $R$ satisfies
\begin{align}
    \mathcal{R}(R)=0, \qquad \mathcal{R}'(R)\geq0,
\end{align}
but a larger root blocks it from the observer. Appendix~\ref{app:mixed-sign} shows that this requires $\kappa > 0$ and four roots (counting multiplicity), of which at least three are positive. The coefficient identity in Equation~\ref{eq:pd-linear-coefficient} excludes four positive roots, forcing the remaining root to be negative. The radial potential is therefore quartic, with roots
\begin{align}
    -u, & \quad x, \quad y,\quad z, \nonumber \\
    u > 0, \qquad & 0 < x \leq y \leq z < r_o. \label{eq:mixed-roots}
\end{align}
This root ordering prevents the photon from crossing $r = 0$. The single negative root and three positive roots have a negative product, and the potential's leading coefficient is positive. The photon potential is therefore negative at zero even though the metric function remains positive there:
\begin{align}
    \mathcal{R}(0)<0, \qquad \Delta_r(0)>0. \label{eq:mixed-zero-signs}
\end{align}
The positive roots can be distinct or repeated, giving three possible orderings:
\begin{enumerate}
    \item Four simple roots, with the inner positive-slope root blocked by the largest simple root.
    \item A simple root satisfying the local derivative test below a larger exact multiplicity-two root with $\mathcal{R}'' > 0$, which is approached asymptotically from the observer side.
    \item An inner exact multiplicity-two root with $\mathcal{R}'' < 0$ below a simple outer root. The potential is negative immediately on either side of this repeated root, so a ray with the same constants of motion cannot approach it from either side.
\end{enumerate}
Every failure of the local derivative rule under these endpoint assumptions therefore has the same mixed-sign root pattern, although different spacetime parameters can produce it.

This root ordering can make the local test fail even when the ray follows a regular path from the observer to its turn. In the example below, NUT charge also keeps $\Sigma$ positive across the entire $r = 0$ surface, removing the Kerr-like curvature singularity.\footnote{In PD, the metric factor $\Sigma = r^2 + (\ell + a\cos\vartheta)^2$ is positive over the entire $r = 0$ surface precisely when $|\ell| > |a|$. In Kerr, $\ell = 0$, and $\Sigma$ vanishes at $r = 0$ on the equator. The same four-simple-root pattern can occur without NUT charge, as in the exact Kerr--Newman--de Sitter example in Appendix~\ref{app:mixed-sign}. That example retains a ring singularity at $r = 0$, while the observed ray follows a regular path.}

As a concrete example, consider a charged, nonrotating NUT--de Sitter spacetime with
\begin{align}
    m&=1, & a&=0, & \alpha_{\rm acc}&=0, \nonumber \\
    \ell&=+\frac{\sqrt5}{2}, & \Lambda&=\frac1{20}, & \beta_{\rm ch}&=\frac{1151}{480}.
\end{align}
The NUT charge keeps $\Sigma = r^2 + 5/4$ positive for every real $r$.\footnote{The Reissner--Nordstr\"{o}m charge bound is modified by NUT charge and $\Lambda$, so $\beta_{\rm ch} > m^2$ alone does not imply a naked singularity.} The radial metric function is
\begin{align}
    \Delta_r(r) = -\frac{r^4}{60} +\frac{7r^2}{8} -2r +\frac{1177}{960},
\end{align}
and since the charge and cosmological constant make $\Delta_r(0) > 0$, the region with $\Delta_r > 0$ extends through zero. For positive $\Lambda$, the coefficient of $r^4$ is negative, so $\Delta_r$ becomes negative at sufficiently large positive and negative radius. The region is therefore bounded by two finite horizons at
\begin{align}
    r_{\rm in} & \approx -8.25344, & r_{\rm out} & \approx 5.84038,
\end{align}
with $\Delta_r > 0$ throughout $(r_{\rm in}, r_{\rm out})$. Pure Taub--NUT instead has $\Delta_r(0) = -\ell^2 < 0$, so its ordinary exterior is separated from $r = 0$ by a region with $\Delta_r < 0$.

In the charged example the ray highlighted in Figure~\ref{fig:nut-counterexample} has $\kappa = 60$, and its radial potential is
\begin{align}
    \mathcal{R}(r) = 2(r+6)(r-1)(r-2)(r-3). \label{eq:nut-example}
\end{align}
All four roots lie between the horizons, and an inward ray from $r_o = 4$ turns at the outermost one, $r = 3$. The potential is negative for $2 < r < 3$, preventing the ray from reaching the inner allowed interval $1 \leq r \leq 2$, even though
\begin{align}
    \mathcal{R}(1) = 0, \qquad \mathcal{R}'(1) = 28 > 0.
\end{align}
All four roots label the same image point algebraically. The positive-derivative condition retains $R=1$ and $R=3$, but only $R=3$ describes the ray's closest approach. The negative root at $r=-6$ makes this global ordering possible even though the observed segment of the ray remains in $3\leq r\leq4$ and never approaches $r=0$ or the negative root.

For this to be a physical counterexample, however, the ray must also be able to reach the observer. We choose the observer inclination and axial impact parameter so that the angular motion is allowed and the ray stays away from both axes:
\begin{align}
    C&=0, & \xi&=8, & \cos\vartheta_o&=-\frac{8\sqrt5}{65}.
\end{align}
For this choice,
\begin{align}
    \Theta(\vartheta_o)=\frac{780}{781}>0, \qquad W>0, \qquad X^2+Y^2<W^2,
\end{align}
so these constants correspond to a real photon direction on the observer's sky, with positive locally measured energy and nonzero radial momentum. Tracing the inward branch backward from $r_o = 4$ takes the ray to its turn at $r = 3$. With its polar motion confined between two turning angles away from both axes, the ray encounters no singularity along this path, and the ray therefore follows a regular path from the observer to $r = 3$, even though the root at $R = 1$ also passes the local slope test. The negative root thus makes this four-simple-root ordering possible in a region that extends through $r = 0$, although this ordering is outside of the conventional positive-radius black hole setting.

The outer turn in this example hides a candidate turn at $R = 1$. An unstable photon orbit at positive radius would have a root of multiplicity two, so hiding it would require two further roots at larger radii (including multiplicity). These four positive roots would contradict Equation~\ref{eq:pd-linear-coefficient} for $m > 0$. A hidden repeated root at positive radius must therefore have $\mathcal{R}'' < 0$, with radial motion forbidden immediately on either side. The broader family considered next permits a hidden unstable orbit when alignment is relaxed.

\subsection{Radial branch selection beyond PD}
\label{sec:op-blocked-orbit}

We now test the limits of the conic construction by considering metrics beyond the PD family. In PD, tidal gravity and the electromagnetic field select the same pair of principal null directions, but the broader type-D Einstein--Maxwell family found by Ovcharenko and Podolsk\'{y} (OP) \cite{ovcharenko_2025_nonaligned,gray_2026_hidden} allows the two fields to select different pairs while preserving the separation of radial and angular photon motion. We use this family to test whether the local radial conditions still identify the turns and unstable photon orbits that an inward ray can reach.

\begin{figure*}[t]
    \centering
    \includegraphics[width=\linewidth]{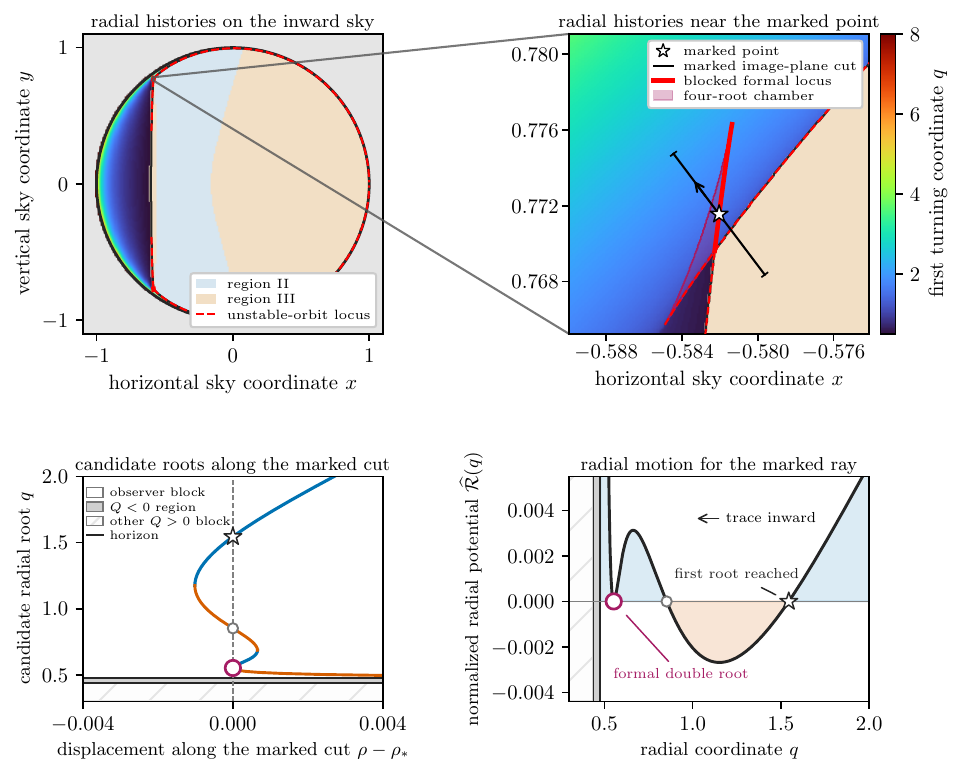}
    \caption{Counterexample showing how an outer turn can hide an unstable photon orbit and break the local derivative test in a particular Ovcharenko--Podolsk\'{y} spacetime. The top left panel shows the inward half of the observer's sky. Color gives the first turning coordinate $q$ for rays that turn before crossing the outermost metric horizon. We find numerically that most inward rays cross this horizon before any turn. Rays in region II cross both positive metric horizons before turning in the next region with $Q > 0$, while those in region III have no real radial root and continue toward the negative-$q$ extension. The top right panel enlarges the neighborhood of the selected image point (star). The red curves correspond to unstable fixed-radius photon orbits: dashed segments trace the physical critical curve, while the thick solid segment marks orbits hidden behind a larger turn. The physical critical curve separates rays that turn outside the horizon from those that cross it and also has segments across which returning rays switch between outer and inner turns. The pink shading identifies image points with four distinct radial roots in the observer's exterior. Along one edge, the middle two roots merge with $\mathcal{R}'' < 0$, so this edge does not belong to the unstable-orbit curve. The bottom-left panel follows the roots along the black image-plane cut, with blue and orange giving positive and negative $\mathcal{R}'$, respectively. Horizontal lines mark metric horizons, and gray and hatched bands show the adjacent regions with $Q < 0$ and $Q > 0$. The bottom-right panel shows $\widehat{\mathcal{R}} = \mathcal{R}/(q^2 + \xi)^2$ for the selected ray. Traced inward from $q_o = 20$, beyond the plotted range, the ray turns at the outermost root. The forbidden interval between the two larger roots prevents it from approaching the unstable orbit.}
    \label{fig:op-critical-counterexample}
\end{figure*}

Figure~\ref{fig:op-critical-counterexample} shows an example of where the local derivative rule fails to select the outermost root in an OP image. In the pink region of the top-right panel, each ray's radial potential has four distinct roots outside the outermost horizon and a positive slope at two of them. Motion is allowed just outside either of these roots and forbidden just inside, so both pass the local test established above for a turn, but the inward ray from the observer reaches only the outermost root. For positive mass, the PD coefficient identity imposed by the field equations forbids this arrangement of four positive roots below the observer. In the OP example, an outer turn can also hide an unstable photon orbit, so the same obstruction affects the construction of the physical critical curve. The solid red segment identifies the image points associated with these hidden orbits.

To follow these rays, we again use the separated radial and angular photon potentials. We obtain them from the OP metric in Carter coordinates $(t, \sigma, p, q)$, where $p$ and $q$ describe angular and radial position. As above, the metric is independent of $t$ and $\sigma$, and its line element can be written as~\cite{ovcharenko_2025_nonaligned,gray_2026_hidden}
\begin{align}
    {\rm d}s^2 = \frac{1}{\Omega^2}\bigg[&-\frac{Q}{p^2 + q^2}({\rm d}t-p^2{\rm d}\sigma)^2 \nonumber\\
    &+\frac{P}{p^2 + q^2}({\rm d}t+q^2{\rm d}\sigma)^2 \nonumber\\
    &+(p^2 + q^2)\left(\frac{{\rm d}q^2}{Q}+\frac{{\rm d}p^2}{P}\right)\bigg].
    \label{eq:op-main-metric}
\end{align}
The quartic functions $P(p)$ and $Q(q)$ describe the angular and radial geometry, and $\Omega(p, q)$ is the conformal factor. Appendix~\ref{app:op-example} gives explicit expressions for $P$, $Q$, and $\Omega^2$ for the solution used below. In these coordinates, $q$ plays the role of $r$ in the preceding sections, labeling radial position (without necessarily measuring proper distance or setting the area of a sphere).

The conserved momenta associated with $t$ and $\sigma$ are $E = -k_t$ and $L = k_\sigma$. For $E > 0$, we write $\xi = L/E$ and $\kappa = K/E^2$, where $K$ is the separation constant. The photon potentials are
\begin{align}
    \mathcal{R}(q) & = (q^2+\xi)^2-\kappa Q(q), \nonumber\\
    \Theta(p) & = \kappa P(p)-(p^2-\xi)^2.
    \label{eq:op-main-separated-potentials}
\end{align}
As in PD, the metric and photon potentials serve different purposes: the roots of $Q$ locate horizons, while the roots of $\mathcal{R}$ locate candidate photon turns. Within a connected region with $Q > 0$, the ray can only propagate where $\mathcal{R} \geq 0$ and $\Theta \geq 0$. At a chosen radius, substituting the photon constants from the local observer frame (Equation~\ref{eq:op-observer-screen}) into $\mathcal{R} = 0$ again gives a quadratic equation on the image. The candidate curves therefore remain conics, with repeated radial roots defining their candidate envelope.

In OP, however, the linear coefficient of $Q$ need not equal $-2m$, so positive mass no longer guarantees the coefficient sign used in Equation~\ref{eq:pd-linear-coefficient}. The candidate conics therefore survive without the root-ordering guarantee that makes the local slope test sufficient in PD. For the image in Figure~\ref{fig:op-critical-counterexample}, we choose a positive-mass OP solution that reduces smoothly to a vacuum PD metric as the electromagnetic field is turned off. Appendix~\ref{app:op-example} gives the exact metric parameters and their definitions for this solution.

We place the observer at $q_o = 20$ and $p_o \simeq -0.32168$ and select the inward-directed ray with the same constants of motion as an unstable photon orbit at $q_\star \simeq 0.55275$. Setting $\xi = -427/2000$ and requiring a root of multiplicity two in the radial potential at this radius fixes $\kappa \simeq 0.938013$. Crucially, the same potential also has two larger simple roots at $q_3 \simeq 0.85241$ and $q_4 \simeq 1.54459$.

These additional roots lie between the observer and the unstable orbit, so the inward ray reaches $q_4$ first. Immediately inside $q_4$, the potential is negative throughout $q_3 < q < q_4$, which forbids further inward motion. The ray therefore turns outward at $q_4$ before it can approach $q_\star$, even though the potential passes the local radial test there. The bottom-right panel of Figure~\ref{fig:op-critical-counterexample} shows this forbidden interval in the (normalized) potential $\widehat{\mathcal{R}} = \mathcal{R}/(q^2 + \xi)^2$. Outer radial barriers also hide unstable photon orbits in rotating wormhole models \cite{gyulchev_2018_wormholeshadow}. Here, the observer and all three photon radii lie in the same positive-radius region outside the outermost metric horizon of a separable Einstein--Maxwell solution.\footnote{In Appendix~\ref{app:op-example}, we show that the selected ray reaches the observer along a regular path from the outer turn.}

In positive-mass PD, the coefficient restriction imposed by the field equations forbids this arrangement of radial roots. The unstable orbit provides a root of exact multiplicity two, and the two larger roots account for the remaining two roots of the quartic potential. We can therefore write
\begin{align*}
    \mathcal{R}(q) = A(q - q_\star)^2(q - q_3)(q - q_4).
\end{align*}
The overall factor $A$ is positive because the potential is positive at the observer, who is positioned beyond all four roots. The two larger roots that block the ray from reaching the unstable orbit bring the total to four positive roots, counting the unstable root twice. With a positive overall multiplier, this arrangement forces the coefficient of $q$ to be negative, contradicting the positive value $2 m \kappa$ required for $m > 0$ and $\kappa > 0$ by the PD coefficient identity in Equation~\ref{eq:pd-linear-coefficient}.

What happens to rays from nearby image points? Small changes in the viewing direction preserve the outer turn, so these rays also turn outward and return along regular paths. The selected image point therefore lies inside the returning region, away from the critical curve. Moving from this point into the pink region splits the inner repeated root in two, giving four distinct roots outside the outermost horizon. The larger of these inner roots has positive slope, just as the outermost root does. Throughout this region, the local test therefore accepts an inner turn that the ray cannot reach.

The bottom-left panel follows these roots as the viewing direction moves along the black image-plane cut. When the inner pair meets while the outer pair remains separate, the resulting unstable orbit stays hidden behind the outer turn, like at the selected image point. When the outer pair meets, the incoming ray approaches the resulting unstable orbit asymptotically. These two possible mergers produce the branches of unstable orbits shown in the top-right panel. The nearly vertical branch follows smaller-radius orbits, while the diagonal branch follows larger-radius orbits. The three dashed segments contribute to the physical critical curve, while the solid red segment marks orbits hidden behind a larger turn.

At the intersection of the two branches, the same image point corresponds to two unstable photon orbits. An inward ray approaches the outer orbit, placing this point on the physical critical curve while the inner orbit remains hidden. The OP example therefore preserves the candidate conics while showing that selecting their physical arcs can still require comparing the radial roots rather than relying on the local derivative test alone. In the positive-radius PD regions considered in Section~\ref{sec:pd-positive-radius}, the coefficient identity excludes these arrangements of roots and makes the local radial test sufficient.

\section{Discussion}
\label{sec:discussion}

We have shown that radius of closest approach defines an astrophysics-independent skeleton for black hole images. For each returning image point covered by the construction we have described, the corresponding light ray's smallest radial coordinate labels that point, and contours of equal depth lie on conics. Separability and screen projection determine the complete candidate conics, while the radial dynamics of the light selects the arcs realized by the rays. On the distant Kerr screen, these arcs tile the initially inward returning image and terminate on the critical curve.

In the more general PD family of spacetimes, the field equations impose a coefficient identity that lets us identify the first radial root reached by an inward ray using a local test of the potential's slope. For positive mass and rays with Killing energy $E > 0$, this test is sufficient on connected positive-radius intervals with $\Delta_r > 0$, radial motion allowed at the inner endpoint, and nonzero inward radial motion at the observer. Without NUT charge, the radial root curves have at most one interior minimum and no interior maximum in the regions of interest. For rays that return or reach an unbounded outer boundary, any finite root on an increasing branch is unique.

In a PD domain that crosses $r = 0$, however, we find that an inward ray can turn before reaching an inner simple root at positive radius, even though that root passes the local test. The PD coefficient identity still prevents an outer turn from hiding an unstable photon orbit at positive radius, though, since any hidden repeated root there must have $\mathcal{R}'' < 0$, with radial motion forbidden immediately on either side. In the OP extension, relaxing the alignment of the principal null directions preserves the conic structure and its formal repeated-root envelope but also allows an outer turn to hide such an orbit.

The same conic family also constrains the critical curve. Eliminating the radius from the repeated-root equations gives a single reflection-symmetric polynomial relation of degree six for generic Kerr and degree eight for generic PD. The physical critical curve is traced by rays that can asymptotically approach an unstable photon orbit along a regular, angularly allowed path. The radius labels thus connect the conic structure of the image to the radial motion of both returning and limiting critical rays.

\subsection{Radial root selection}

For an initially inward returning ray, the closest-approach map
\begin{align}
    (\alpha,\beta)\longmapsto T(\alpha,\beta)
\end{align}
records the smallest radial coordinate reached along its path. This limits where the light at that image point can originate: if the emitting material is confined to $r < R$, a ray with $T > R$ cannot intersect it, but a ray with $T < R$ enters that radial region (although its intensity still depends on the distribution of emitting material and on radiative transfer along the entire history of its path). The contours therefore identify which returning rays can sample a chosen radial region before an emission model is specified.

The positive-radius PD result in Section~\ref{sec:pd-positive-radius} lets us construct the map one contour at a time. Along the candidate conic at each radius $R$, the local slope test identifies image points whose first radial root is $R$. Applying the angular and regular-path conditions then gives the physical contour, without requiring a manual comparison between the radial roots at every image point.

\subsection{Radial structure of the Kerr ``D shape''}

\begin{figure*}[t]
    \centering
    \includegraphics[width=\linewidth]{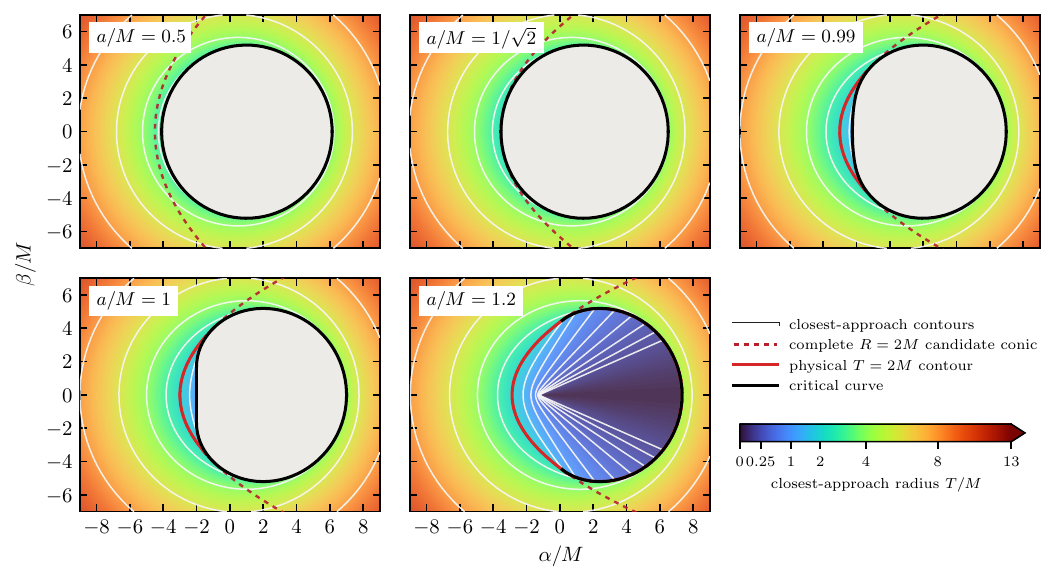}
    \caption{Closest-approach maps for a distant equatorial observer in Kerr as spin increases. In the black hole panels with $a/M \leq 1$, color encodes the outer radial turning radius $T/M$, and white curves connect image points with the same closest approach. The dashed red curve is the complete parabolic candidate conic at $R = 2M$, the solid red arc is its physical $T = 2M$ contour, and the black curve is the critical curve. At $a/M = 1/\sqrt{2}$, the prograde critical radius first reaches $2M$. At higher spin, returning rays can turn at $R < 2M$, and their hyperbolic contours occupy an increasing portion of the prograde image. At $a = M$, the vertical prograde segment of the critical curve is the near-horizon extremal Kerr (NHEK) line. The $a/M = 1.2$ panel colors the largest positive root in the formal continuation to horizonless superextremal Kerr.}
    \label{fig:kerr-shapes}
\end{figure*}

The closest-approach map shows how spin changes the radial depths sampled on opposite sides of a Kerr image \cite{bardeen_1973_kerrgeo,teo_2003_photonskerr}. In Schwarzschild, all critical rays approach the same radius $R = 3M$, but increasing spin separates the radii of the prograde and retrograde spherical orbits, allowing prograde rays to approach more closely while retrograde rays turn farther out. For an equatorial observer, the limiting prograde and retrograde radii approach $M$ and $4M$, respectively, as $a \to M$.

For an equatorial observer, the $R = 2M$ conic is a parabola for every nonzero spin, so the red curve in Figure~\ref{fig:kerr-shapes} provides a fixed reference as the prograde turning radius moves inward. The conics at larger radii are ellipses, while those at smaller radii are hyperbolas. At $a/M = 1/\sqrt{2}$, the prograde critical radius reaches $2M$, and returning rays can turn at $R < 2M$ as the spin increases further. Their closest-approach contours then include hyperbolic arcs, which occupy an increasing part of the prograde image.

Figure~\ref{fig:kerr-tiling} shows nearly circular tilings for both spins when viewed close to the rotation axis. At higher inclination, the small-radius contours become more displaced and their endpoints locate the prograde part of the critical curve. In the right-most panel, the arcs corresponding to the smallest radii are steep, and their endpoints trace the nearly vertical prograde edge. The contours for larger radii are more ellipse-like and meet the rounded retrograde side. The familiar D shape thus records both the changing conic geometry and the different radial depths probed on opposite sides of the image.

For sufficiently inclined observers, the smallest prograde turning radii approach the horizon as $a \to M$. At extremality, the root and derivative conics at $R = M$ share a line, and the segment selected by radial accessibility and angular visibility is the vertical near-horizon extremal Kerr (NHEK) line. Appendix~\ref{app:kerr-extensions} gives the inclination condition for this segment to be visible.

The $a/M = 1.2$ panel continues the map to horizonless Kerr, with each image point labeled by the largest positive radial root. The hyperbolic contours now extend across the central image, where the black hole panels contain a shadow. The increasing extent of these contours (i.e., their ``hyperbolitude'') suggests a possible image signature of horizonless Kerr models, e.g., if emission from small radii illuminates arcs within this region.

\subsection{The critical curve as a constrained conic envelope}
\label{sec:critical-curve-restrictions}

How much freedom does the algebra of the conic construction allow in the shape of the critical curve? In general, a freely chosen family of conics can have any smooth, regular reflection-symmetric curve $Y^2 = g(X)$ as its envelope over a finite closed interval of $X$ (Appendix~\ref{app:critical-curve-restrictions}). In Kerr and PD, however, the conics at different radii all come from the same radial potential and the same observer. Their coefficients are therefore linked, and these relations lead to a single polynomial equation for the candidate critical curve.

The equation can be found by eliminating the orbit radius from the repeated-root conditions $\mathcal{C}_0(X, Y; R) = \mathcal{C}_1(X, Y; R) = 0$. A repeated root requires the radial discriminant to vanish, giving a polynomial relation among the constants of motion alone. For the PD potential $\mathcal{R} = (r^2 + c)^2 - \kappa\Delta_r$, with $c = a^2 + \ell^2 - 2a\ell C - a\xi$, we denote the discriminant as $\mathcal{H}_\Delta(c, \kappa)$, after removing its universal $\kappa^2$ factor.\footnote{At $\kappa = 0$, the radial potential is a perfect square, giving the universal factor in Equation~\ref{eq:reduced-radial-discriminant}. On the real adapted image, this branch has no positive-radius critical point.}

To express this repeated-root condition on the image, we substitute the photon constants associated with each image point. Using Equation~\ref{eq:pd-screen-map}, we write $c = c_o - hX$ and $\kappa = q_o (X^2 + Y^2)$, where $c_o = (\ell + a \cos\vartheta_o)^2$, $h = a p_1$. The coefficients $c_o, h$, and $q_o$ are fixed by the spacetime and observer. The condition $H_\Delta(c, \kappa) = 0$ therefore becomes
\begin{align}
    \mathcal{H}_\Delta \left(c_o - h X, q_o (X^2 + Y^2)\right) = 0. \label{eq:pd-critical-octic}
\end{align}
This is a single polynomial equation in the image coordinates, with the orbit radius eliminated. Every repeated-root image point with $\kappa \ne 0$ satisfies it. For generic PD, the polynomial has total degree eight and contains only even powers of $Y$, giving reflection symmetry across the horizontal image axis. For generic subextremal Kerr, it reduces to the sixth-degree relation derived by \citet{gralla_2020_observableshape}.\footnote{Schwarzschild, extremal Kerr, nonrotating or face-on screens, repeated horizons, and drops in radial polynomial degree lead to additional factors or lower-degree limits. We discuss this point in greater depth in Appendix~\ref{app:critical-curve-restrictions}.}

These equations restrict which critical curves can appear in Kerr and PD. For a candidate point to lie on the physical critical curve, its ray traced backward from the observer must asymptotically approach an unstable photon orbit along a regular, angularly allowed path. The critical curve is therefore selected from the polynomial curve by the same propagation conditions that determine which conic arcs belong to the returning image.

\subsection{Scope of the black hole interpretation}

The positive-radius result described in Section~\ref{sec:pd-positive-radius} applies directly to initially inward rays with $E > 0$ in standard asymptotically flat Kerr--Newman and nonaccelerating Kerr--Newman--AdS black holes with positive mass. In these cases, the event horizon provides an inner boundary where radial motion is allowed, so an inward ray with nonzero radial momentum at a finite observer satisfies the endpoint conditions for the local derivative test. For a distant image in the limit $r_o \to \infty$, radial motion must also remain allowed at sufficiently large radius.

For an accelerating black hole, the image also depends on which exterior region contains the observer and the location of its outer boundary. The conformal boundary $\Omega = 0$ generally depends on angle, so constructing a distant image requires specifying where the observer approaches it and how the image coordinates are scaled (Appendix~\ref{app:boundary-screens}). The observer and rays must remain in a regular angular sector and in the same connected region with $\Omega > 0$, outside the event horizon. The physical contours then consist of rays whose angular motion is allowed and whose paths remain regular in this region.

Once the spacetime, observer, and image projection are specified, the same depth map can be used to compare images from different emission models. The contours identify which returning rays can enter a chosen radial region, so they show where emission from that region can appear in the returning image. For example, in the Kerr images above, returning rays on the prograde side can encounter material at radii that their retrograde counterparts cannot reach. Emission confined to these small radii can therefore contribute to only part of the returning image, while a source extending to larger radii can illuminate rays that turn farther out. The plasma properties and radiative transfer determine how bright these regions appear, but the radial restrictions are common to all such models. Comparing synthetic images with the depth map can then identify which features could contain light from the innermost flow and help interpret how their appearance changes with the emission model.

\begin{acknowledgments}
G.N.W.~gratefully acknowledges support from the Institute for Advanced Study, through the Taplin Fellowship, and from the Leinweber Forum for Theoretical Physics. A.M.~acknowledges support from the Ambrose Monell Foundation, the W.M.~Keck Foundation and the John N.~Bahcall Fellowship Fund at the Institute for Advanced Study. Generative AI tools including ChatGPT and Claude were used for language editing and exploratory checks of algebraic manipulations. The authors independently verified all arguments and calculations and take full responsibility for the manuscript.
\end{acknowledgments}

\appendix

\section{Kerr extensions}
\label{app:kerr-extensions}

Here, we extend the Kerr construction provided in Section~\ref{sec:kerr} to extremal spin, take its Schwarzschild limit, and describe how observer motion changes the image.

\subsection{Extremal Kerr and the horizon component}

At $a = M$, the critical curve can include a segment whose rays approach the horizon. We first show that exterior turning roots still obey the branch rule used for subextremal Kerr, then identify the additional horizon component. The case $a = -M$ follows by reflecting $(a, \xi, \alpha) \mapsto (-a, -\xi, -\alpha)$ while leaving $(\eta, \kappa, \beta)$ unchanged. To examine the exterior radial root curve, set
\begin{align}
    x&=r-M>0, & d&=M(2M-\xi).
\end{align}
The functions that determine its zeros and stationary points become
\begin{align}
    \Delta &= x^2, & P_\xi&=x^2+2Mx+d, \nonumber \\
    G_\xi &= x(x^2-d), & F_\xi&=\left(x+2M+\frac{d}{x}\right)^2. \label{eq:extremal-kerr-one-well}
\end{align}
If $d < 0$, $P_\xi$ has one positive zero while $G_\xi > 0$, so $F_\xi$ decreases to zero and then increases. If $d > 0$, $P_\xi > 0$ and $G_\xi$ changes sign only at $x = \sqrt d$, again giving one minimum. For $d = 0$, $F_\xi = (x+2M)^2$ is strictly increasing. Every exterior turning root $R > M$ therefore obeys the same derivative branch rule as in the subextremal case.

An unstable spherical photon orbit outside the horizon satisfies $\mathcal{R} = \mathcal{R}' = 0$ and $\mathcal{R}'' > 0$. Solving these conditions at $a = M$ gives
\begin{align}
    \xi_c(R)&=M+2R-\frac{R^2}{M}, & \kappa_c(R)&=4R^2, \nonumber \\
    \eta_c(R)&=\frac{R^3(4M-R)}{M^2}, & \mathcal{R}''(R)&=8R(R-M)>0. \label{eq:extremal-kerr-outer-branch}
\end{align}
For every $R > M$, Equation~\ref{eq:extremal-kerr-outer-branch} shows that the root has multiplicity two. As $R \downarrow M$, the orbit constants approach $(\xi, \kappa, \eta) = (2M, 4M^2, 3M^2)$, where the horizon root has multiplicity three. To examine other rays with the same limiting impact parameter, set $\xi = 2M$ in the full radial potential:
\begin{align}
    \mathcal{R}(r) =(r-M)^2\left[(r+M)^2-\kappa\right]. \label{eq:extremal-kerr-horizon-factorization}
\end{align}
For $\kappa < 4M^2$, a ray on this branch approaches the horizon asymptotically and has no exterior turning root. At $\kappa = 4 M^2$ the horizon root is triple and joins Equation~\ref{eq:extremal-kerr-outer-branch}; for $\kappa > 4M^2$, an outer root at $r = \sqrt\kappa-M$ blocks access to the horizon.

On the image, the horizon component follows from the same two conics used to construct the exterior critical curve. At $R = M$, they become
\begin{align}
    \mathcal{C}_0&=M^2(2M-\xi)^2, & \mathcal{C}_1&=4M^2(2M-\xi), \label{eq:extremal-kerr-degenerate-conics}
\end{align}
so both vanish along the line $\xi = 2M$. The corresponding coordinates on a distant screen satisfy
\begin{align}
    \alpha_{\rm H}&=-2M\csc\theta_o, \nonumber \\
    \beta^2&=\kappa-M^2\sin^2\theta_o -4M^2\cot^2\theta_o.
\end{align}
A ray can approach the horizon only if no larger root blocks its path, which requires $\kappa \leq 4 M^2$. Substitution gives the vertical extent of the horizon component,
\begin{align}
    |\beta| \leq M\sqrt{3+\cos^2\theta_o-4\cot^2\theta_o}, \label{eq:extremal-kerr-nhekline}
\end{align}
provided the expression under the square root is nonnegative, or equivalently $\sin\theta_o \geq \sqrt{3} - 1$. For strict inequality, this is the vertical NHEK line segment of the extremal critical curve, whose endpoints are the $R \downarrow M$ limits of the exterior spherical-orbit branches \cite{gralla_2020_observableshape}. At equality, the segment collapses to a point.

\subsection{Schwarzschild shared-circle limit}

In Schwarzschild, all critical rays approach the same photon-orbit radius, so varying $R$ no longer traces the critical curve. To see this, set $a = 0$ and let $\rho^2 = \alpha^2+\beta^2$. The fixed-root and derivative conics reduce to concentric circles,
\begin{align}
    \mathcal{C}_0 &= R^4 - R (R - 2M) \rho^2 = 0, \nonumber \\
    \mathcal{C}_1 &= 4R^3 - 2(R - M) \rho^2 = 0.
\end{align}
For a generic exterior radius $R > 2 M$, these circles are distinct and do not intersect. At $R = 3M$, however, both equations become
\begin{align}
    \rho^2 = 27M^2.
\end{align}
This shared circle is the familiar Schwarzschild critical curve, and its finite radius shows how the physical image remains regular as $a \to 0$, even though the rotating parametrization in Equations~\ref{eq:kerr-xi-critical}--\ref{eq:kerr-critical-parametrization} becomes singular.

\subsection{Aberration and image projection}
\label{app:abbreation}

Changing the observer's velocity changes the apparent direction of each photon and therefore changes the apparent locations of the closest approach contours on the sky. Let $(x, y, z)$ be the components of the photon's unit spatial direction in the observer's orthonormal frame, and define the stereographic sky coordinate
\begin{align}
    w=\frac{x+iy}{1+z}.
\end{align}
Lorentz aberration acts on the celestial sphere by a M\"{o}bius transformation,
\begin{align}
    w'=\frac{A_Lw+B_L}{C_Lw+D_L}, \qquad A_LD_L-B_LC_L\neq0.
\end{align}
Each ray retains its turning radius, so aberration carries the returning region and critical curve to the moving observer's sky while preserving the radius labels, but whether the contours appear as plane conics depends on the image projection. An induced projective map preserves their conic form, as in Equation~\ref{eq:kerr-projective-camera}, but a generic stereographic projection can turn them into nonconic curves.

An observer with any timelike velocity can construct an image with the same conic contours by correcting the measured photon directions for aberration. This correction recovers the directions in the ZAMO frame of Section~\ref{sec:kerr-finite-observers} for Kerr or GPL's frame in Equation~\ref{eq:pd-observer-tetrad} for PD, at any position where that frame is defined. Applying the corresponding orthographic projection then gives the conic image. For a specified spacetime and observer, the change of frame and projection are fixed and apply to every ray and turning radius.

\section{Proof of the positive-radius branch rule}
\label{app:pd-branch-proof}

An inward ray only encounters the largest radial root below the observer. We show that, under the positive-radius and endpoint assumptions of Section~\ref{sec:pd-positive-radius}, every root that passes the local slope test is precisely this first-encountered root. In practice, our proof counts the roots that would be needed to hide an inner root passing the test, then uses the PD coefficient identity to rule out that arrangement.

At any interior root, $\Delta_r > 0$ gives
\begin{align}
    \kappa = \frac{(R^2+c_\xi)^2}{\Delta_r(R)} \geq 0. \label{eq:pd-root-kappa}
\end{align}
If $\kappa = 0$, then
\begin{align}
    \mathcal{R}(r)=(r^2+c_\xi)^2.
\end{align}
This polynomial has no positive root when $c_\xi \geq 0$ and exactly one positive root when $c_\xi < 0$. In the latter case, the root has multiplicity two and $\mathcal{R}''(R) = 8R^2 > 0$. Since there is no larger positive root to block it, Equation~\ref{eq:pd-positive-branch} holds for $\kappa = 0$.

Now let $\kappa > 0$. Since $m > 0$, the PD field equations require the coefficient of $r$ in the potential to be positive:
\begin{align}
    [r]\mathcal{R} = 2m\kappa>0. \label{eq:pd-linear-coefficient-strict}
\end{align}
Suppose an interior root $R$ passes the test $\mathcal{R}'(R) \geq 0$ but a larger root blocks access from the observer, and let $S > R$ be the largest root below the observer. Since $\mathcal{R}(r_o) > 0$ and there are no roots between $S$ and $r_o$, the potential is positive immediately to the right of $S$.

If $R$ is simple, then $\mathcal{R}'(R) > 0$, so the potential is negative immediately to its left and positive immediately to its right. To recover the allowed radial motion at the inner endpoint, another root must lie in $[r_{\rm in}, R)$. If $S$ has even multiplicity, it counts at least twice, bringing the total to at least four positive roots. If $S$ has odd multiplicity, the potential is negative immediately to its left, so a further sign-changing root is needed between $R$ and $S$. The total is again at least four, counting multiplicity.

If $R$ is repeated, it already counts at least twice. When the potential is positive immediately to its right, an even-multiplicity $S$ brings the count to four, while an odd-multiplicity $S$ requires an additional sign-changing root between $R$ and $S$. When the potential is negative immediately to the right of $R$, there are two possibilities. If $R$ is of even multiplicity, it will also have a negative potential immediately to its left, requiring an earlier root to restore the allowed motion at the inner endpoint. Alternatively, an odd-multiplicity $R$ has multiplicity at least three. Including $S$ therefore gives at least four positive roots in every case.

A root hidden from the observer can therefore pass the local test only if the potential has at least four positive roots (counted with multiplicity). This is impossible when $\mathrm{deg} \, \mathcal{R} \leq 3$. For a quartic, these roots account for the entire polynomial, so we can write
\begin{align}
    \mathcal{R}(r) =A\prod_{i=1}^4(r-r_i), \qquad r_i>0,
\end{align}
observer-side positivity implies $A > 0$, and Vieta's formula gives
\begin{align}
    [r]\mathcal{R} =-A\sum_{i<j<k}r_ir_jr_k<0.
\end{align}
This contradicts the positive coefficient in Equation~\ref{eq:pd-linear-coefficient-strict}, and every root satisfying $\mathcal{R}'(R) \geq 0$ is therefore the first root approached by an inward ray from the observer.

The first root also passes the local test, completing the converse. Let $T$ be the largest root below the observer. Since $\mathcal{R} > 0$ between $T$ and $r_o$, a simple root at $T$ has $\mathcal{R}'(T) > 0$, while a repeated root has $\mathcal{R}'(T) = 0$. If the repeated root had multiplicity three, the potential would be negative immediately to its left, requiring an earlier positive root by the inner-endpoint condition. A root of multiplicity four would itself account for all four roots. Both arrangements give the same Vieta contradiction, so the only possible repeated outer root has exact multiplicity two. Observer-side positivity then gives
\begin{align}
    \mathcal{R}''(T)>0.
\end{align}

The multiplicity of the selected radial root determines whether the ray turns in finite affine parameter or approaches the selected radius asymptotically. To see this directly, introduce the energy-normalized Mino parameter $\tau$ by
\begin{align}
    \frac{d\tau}{ds} =\frac{E\Omega^2}{\Sigma}, \qquad \left(\frac{dr}{d\tau}\right)^2 =\mathcal{R}(r),
\end{align}
where $s$ is affine parameter increasing along the future-directed photon on the $E>0$ branch. Near a selected root of multiplicity $q$,
\begin{align}
    \mathcal{R}(r) \sim A_q(r-T)^q, \qquad A_q>0
\end{align}
on the observer side. Near the root, the Mino parameter required to reach $T$ is proportional to the integral of $(r-T)^{-q/2}$, which is finite for a simple root ($q = 1$) and divergent for a root of multiplicity two ($q = 2$). The same distinction holds for affine parameter when $\Sigma/(E\Omega^2)$ is positive, bounded above, and bounded away from zero along the observer-to-root interval, so that converting between the two parameters preserves whether the parameter interval is finite or infinite. Under these conditions, the ray reaches a simple root and turns in finite affine parameter, while it approaches a repeated root asymptotically.

If radial motion remains allowed at sufficiently large radius in an unbounded interval, one can simply choose a finite auxiliary outer endpoint beyond every finite root. The proof then applies on the interval between that endpoint and $r_{\rm in}$, establishing Equation~\ref{eq:pd-unbounded-branch} by the same finite-radius argument.

\section{Radial root structure without NUT charge}
\label{app:no-nut}

When the NUT charge vanishes, the shape of the radial root curve provides another way to identify the first turn reached by an inward ray. We prove the result of Section~\ref{sec:no-nut-one-well}: for positive mass, the curve has at most one interior minimum and no interior maximum in each of the specified outer intervals. Set $\ell = 0$ and write
\begin{align}
    P_c(r)&=r^2+c, & F_c(r)&=\frac{P_c(r)^2}{\Delta_r(r)}, \label{eq:no-nut-Fc}
\end{align}
where $c = a^2 - a\xi$ is fixed by the photon impact parameter. On an interval with $\Delta_r > 0$, a horizontal line at height $\kappa$ intersects $F_c$ at the radial roots. At each intersection,
\begin{align}
    \mathcal{R}'(R)=\Delta_r(R)F_c'(R), \label{eq:no-nut-slope-identity}
\end{align}
so the increasing side of $F_c$ selects precisely the roots with positive radial slope. To determine how many such roots can occur, we locate the minima and maxima of $F_c$. Zeros of $P_c$ give minima with $F_c = 0$. Away from these zeros, the condition $F_c' = 0$ can be solved for the impact-parameter combination $c$:
\begin{align}
    c=C_\Delta(r), \qquad C_\Delta(r) =\frac{4r\Delta_r}{\Delta_r'}-r^2. \label{eq:no-nut-Cdelta}
\end{align}
At such a point, $F_c'' = 4r C_\Delta'/\Delta_r$, so a positive $C_\Delta'$ gives a minimum. We therefore examine how $C_\Delta$ varies across each radial interval. Its derivative is conveniently expressed using the logarithmic derivative of $\Delta_r$ and its negative derivative,
\begin{align}
    L=\frac{\Delta_r'}{\Delta_r}, \qquad Z=-L'.
\end{align}
Where $\Delta_r' \neq 0$, differentiation gives
\begin{align}
    C_\Delta'(r) =\frac{2}{L^2} \left(2L + 2rZ - rL^2\right). \label{eq:no-nut-Cdelta-prime}
\end{align}
The sign of $C_\Delta'$ is set by the factor $2L + 2rZ - rL^2$. We now show that it is positive throughout each outer interval, then use the endpoint values to determine whether $F_c$ has a minimum or is monotonic. The metric function can be factored according to the sign of
\begin{align}
    q = \alpha_{\rm acc}^2 + \frac{\Lambda}{3}
\end{align}
which fixes the coefficient of $r^4$ in $\Delta_r$. Together with the coefficient signs in Equation~\ref{eq:no-nut-coefficients}, this gives the four cases described below.

\subsection{Finite interval between horizons}

For $q > 0$, the radial interval of interest lies between the two outermost distinct positive horizons. Begin by factoring the radial metric function as
\begin{align}
    \Delta_r =-K_4\prod_{i=1}^4(r-r_i), \qquad K_4>0, \label{eq:no-nut-finite-factor}
\end{align}
with
\begin{align}
    r_1\leq0\leq r_2\leq r_3<r_4, \qquad I=(r_3,r_4), \qquad r_3>0.
\end{align}
The logarithmic derivatives can now be written in terms of the distances to the four roots. For $r \in I$, set
\begin{align}
    x &= \frac{r}{r-r_1}, & y &= \frac{r}{r-r_2}, & z &= \frac{r}{r-r_3}, & t &= \frac{r}{r_4-r}.
\end{align}
Then $0 < x \leq 1$, $y \geq 1$, $z > 1$, and $t > 0$. With $p = x+y+z$,
\begin{align}
    r\left(2L + 2rZ - rL^2\right) &= t^2 + 2(p-1)t \nonumber \\
    &\quad +2(x^2+y^2+z^2) \nonumber \\
    &\quad -p^2+2p.
\end{align}
Writing $s = y+z$ and using $2(y^2+z^2) \geq s^2$ bounds the last line below by
\begin{align}
    x^2+2x+2s(1-x)>0.
\end{align}
This bound gives $C_\Delta' > 0$ on both sides of the radius $r_m$ where $\Delta_r$ reaches its unique maximum. Between the inner horizon and $r_m$, and between $r_m$ and the outer horizon, the ranges of $C_\Delta$ are
\begin{align}
    C_\Delta:(r_3,r_m)&\longrightarrow(-r_3^2,\infty), \nonumber \\
    C_\Delta:(r_m,r_4)&\longrightarrow(-\infty,-r_4^2). \label{eq:no-nut-finite-C-ranges}
\end{align}
These ranges do not overlap, so a fixed $c$ gives at most one stationary point with $P_c \neq 0$. Such a point is a minimum because $C_\Delta' > 0$. For $-r_4^2 < c < -r_3^2$, there is instead one zero of $P_c$ inside the interval, giving the unique minimum $F_c = 0$. The curve therefore has at most one interior minimum.

When $P_c$ vanishes at a horizon, the numerator and denominator of $F_c$ vanish together, so we evaluate that limit separately. If $c = -r_3^2$ and $r_3$ is simple, $F_c$ increases from zero at the inner endpoint to infinity at $r_4$. If $c = -r_4^2$, it decreases from infinity to zero at the outer endpoint, where radial motion is forbidden nearby for $\kappa > 0$. If the inner horizon has multiplicity two, $r_2 = r_3 = R$, then $y = z$ and the same bound still gives $C_\Delta' > 0$. At $c = -R^2$, canceling the common factors gives
\begin{align}
    F_c(r) =\frac{(r+R)^2}{K_4(r-r_1)(r_4-r)},
\end{align}
which increases from a finite horizon value to infinity. A root of even multiplicity leaves the sign of $\Delta_r$ unchanged across it, so it cannot provide the outer boundary of the finite positive interval considered here.

\subsection{Vanishing quartic coefficient}

When $q = 0$, the quartic term vanishes and the remaining degree depends on the acceleration. First take $\alpha_{\rm acc} = 0 = \Lambda$, for which $\Delta_r$ is quadratic with positive leading coefficient $K_2$:
\begin{align}
    \Delta_r = K_2 (r - r_-) (r - r_+), \qquad 0 \leq r_- \leq r_+,
\end{align}
The exterior interval therefore extends from the larger horizon at $r_+$ to infinity, $I = (r_+, \infty)$. For a simple outer horizon, set $x = r/(r-r_-)$ and $y = r/(r-r_+)$. Both are at least one, and the factor controlling $C_\Delta'$ is
\begin{align}
    r\left(2L + 2rZ - rL^2\right) = (x - y)^2 + 2x + 2y > 0.
\end{align}
The horizon value and large-radius limit of $C_\Delta$ are
\begin{align}
    C_\Delta(r_+)=-r_+^2, \qquad C_\Delta(r)\longrightarrow\infty \quad(r\to\infty).
\end{align}
Since $C_\Delta$ increases between these endpoints, $F_c$ has one exterior minimum with $P_c \neq 0$ when $c > -r_+^2$. For $c < -r_+^2$, the numerator vanishes once, again giving a single minimum. At $c = -r_+^2$, the numerator and denominator vanish at the horizon and $F_c$ is strictly increasing outside it. If the horizons are at the same location $r_- = r_+ = h$, then
\begin{align}
    C_\Delta(r)=(r-h)^2-h^2,
\end{align}
and canceling the common horizon factors gives $F_{-h^2} = (r+h)^2/K_2$. Both expressions increase for $r > h$, so the classification also includes the extremal quadratic exterior.

When $q = 0$ with $\alpha_{\rm acc} \neq 0$, canceling the quartic term leaves a cubic with positive leading coefficient. The coefficient signs and the existence of the exterior interval give
\begin{align}
    \Delta_r &=K_3(r-r_1)(r-r_2)(r-r_3), \nonumber \\
    r_1&<0\leq r_2\leq r_3=R_h,
\end{align}
where $r_2 = 0$ is included when $b_0 = 0$. To evaluate the same derivative factor on $r > R_h$, define
\begin{align}
    x &= \frac{r}{r-r_1}, \\
    y &= \frac{r}{r-r_2}, \\ 
    z &= \frac{r}{r-r_3}.
\end{align}
Then $0 < x < 1$, $y \geq 1$, and $z > 1$. With $p = x+y+z$, $s = y+z$, and $Z = (x^2+y^2+z^2)/r^2$, then
\begin{align}
    r \left(2L + 2rZ - r L^2\right) \geq x^2 + 2x + 2s (1-x) > 0.
\end{align}
The same monotonicity argument gives at most one exterior minimum of $F_c$. If the largest horizon root has multiplicity two, $r_2 = r_3 = R_h$, the inequality remains strict with $y = z$. When the numerator also vanishes at this horizon, canceling the common factors gives
\begin{align}
    F_{-R_h^2}(r) &=\frac{(r+R_h)^2}{K_3(r-r_1)}, \nonumber \\
    F_{-R_h^2}'(r) &=\frac{(r+R_h)(r-2r_1-R_h)} {K_3(r-r_1)^2}>0.
\end{align}
Thus $F_c$ is increasing in this limit as well. A positive horizon of multiplicity three would contradict $b_0 \geq 0$, so these cases cover the coincident horizons allowed for the cubic.

\subsection{AdS-like quartic exterior}

For $q < 0$, the metric function grows as a positive quartic and the selected radial interval extends from the largest positive horizon to arbitrarily large radius. To prove that $C_\Delta' > 0$ throughout this exterior, set $\mu = -q > 0$ and factor $\Delta_r$ as
\begin{align}
    \Delta_r =\mu(r-C)(r-R_h)(r^2+Sr+P), \label{eq:no-nut-ads-factor}
\end{align}
where
\begin{align}
    0 & \leq C\leq R_h, & S & \geq C+R_h, \nonumber \\
    P(C + R_h) & > SCR_h, & P & > 0. \label{eq:no-nut-ads-conditions}
\end{align}
The endpoint $C = 0$ includes $b_0 = 0$, while $C = R_h$ gives a horizon with multiplicity two. The remaining quadratic factor introduces two additional coefficients, so we bound $2L + 2rZ - rL^2$ over their allowed range.

To find a bound that is valid throughout the exterior, fix an arbitrary radius $r > R_h$ and let the quadratic coefficients $S$ and $P$ vary over their allowed range. We express the constraints on these coefficients in dimensionless form by setting
\begin{align}
    c &= \frac{C}{r}, & d &= \frac{R_h}{r}, \nonumber \\
    U &= c+d, & V &= cd, \nonumber \\
    s &= \frac{S}{r}, & p&=\frac{P}{r^2}.
\end{align}
The coefficient bounds then become $0 \leq c \leq d < 1$, $s \geq U$, and $p > Vs/U$. To collect the contributions from the two horizon factors and the remaining quadratic, we introduce
\begin{align}
    m &= \frac{2 - U}{1 - U + V}, & w &= \frac{1}{1 - U + V}, \nonumber \\
    n &= \frac{s + 2}{1 + s + p}, & v &= \frac{1}{1 + s + p}.
\end{align}
In these dimensionless variables, the derivative factor becomes
\begin{align}
    & r\left(2L+2rZ-rL^2\right) = \mathcal F, \nonumber \\
    & \mathcal F = (m-n)^2+2(m+n)-4(w+v). \label{eq:no-nut-ads-F}
\end{align}

We can bound this expression by first varying $p$ at fixed $s$, then varying $s$ along the lower boundary of the allowed coefficient region. At fixed $s$,
\begin{align}
    \frac{\partial\mathcal F}{\partial p} = \frac{2}{(1+s+p)^2} \left[(s+2)(m-1)+2- \frac{(s+2)^2}{1+s+p}\right].
\end{align}
The expression in brackets increases with $p$, so it is enough to evaluate it at the lower boundary $p = Vs/U$. At that boundary, the bracketed expression has the same sign as the polynomial $E = A_2s^2+A_1s+A_0$, whose coefficients satisfy
\begin{align}
    A_0 &= 2(U-2V) > 0, \nonumber \\
    UA_2 &= c^2(1-d)+d^2(1-c) + c\,d\,(3-c-d-cd)>0, \nonumber \\
    UA_1 &= c\,(1-d)+d\,(1-c) +2c^2(1-d)+2d^2(1-c) \nonumber \\
    &\quad + 5\,c\,d\,(2-c-d)>0.
\end{align}
All three coefficients are positive, so $\mathcal{F}$ increases strictly with $p$, and we can bound it from below by setting $p = Vs/U$ at each fixed $s$.

On that boundary, set $G = 1+V/U$. Then
\begin{align}
    \frac{d\mathcal F}{ds} =\frac{2}{(1+sG)^2} \left[1+(2G-1)(m-n)\right],
\end{align}
and
\begin{align}
    \mathcal D &= U\,(1-U+V)(1+sG), \nonumber \\
    \mathcal D(m-n) &= U\,(U-2V) +s(U+2V-2UV)>0.
\end{align}
These expressions show that $\mathcal{F}$ also increases with $s$ along this boundary. Since $s \geq U$, its lowest boundary value occurs at $s = U$ and $p = V$, where
\begin{align}
    \mathcal F_{\rm boundary} =\frac{4\,(d^2-c^2)^2} {(1-c^2)^2(1-d^2)^2}\geq0.
\end{align}
This boundary value is positive for $c < d$, including when $c = 0$, and vanishes only when the two horizon radii coincide at $c = d$. In that case, the physical coefficients still satisfy $p > Vs/U$ because $b_1 < 0$, so strict increase with $p$ gives $\mathcal{F} > 0$. Hence $C_\Delta$ increases throughout every admitted AdS-like exterior, including the zero-root and coincident-horizon limits.

For rays arriving from arbitrarily large radius, the branch rule requires radial motion to remain allowed in the outer region. We therefore need to determine which photon constants satisfy this endpoint condition before selecting the first radial root. At large $r$, the radial root curve $F_c$ and photon potential $\mathcal{R}$ behave as
\begin{align}
    F_c(r)\longrightarrow\frac1\mu, \qquad \qquad \mathcal{R}(r) \sim(1-\mu\kappa)\,r^4.
\end{align}
Radial motion at sufficiently large radius is therefore allowed for
\begin{align}
    \kappa < \frac{1}{\mu}.
\end{align}
When $\kappa = 1/\mu$, the quartic term vanishes, so the remaining terms determine whether a photon can reach arbitrarily large radius. For nonzero acceleration, the cubic term is negative and blocks access to large radius. When acceleration vanishes, the quadratic and lower-order terms determine whether radial motion remains allowed. For $\kappa < 1/\mu$, any root on the increasing side of $F_c$ is the first finite root reached by an inward ray from large radius.

In all four cases, $F_c$ is monotonic or has a single interior minimum, including when horizons coincide or the numerator and denominator vanish at the same radius. The endpoint limits determine which levels $F_c = \kappa$ correspond to rays that return or reach an unbounded outer boundary. For these levels, any finite root on the increasing side is unique and is the first root approached from larger radius. This establishes the radial classification in Section~\ref{sec:no-nut-one-well}. To construct a black hole image, the selected interval must also contain the required regular photon paths outside an event horizon; for accelerating solutions, the observer and outer boundary must be specified as in Section~\ref{sec:discussion}.

\section{Radial intervals crossing zero}
\label{app:mixed-sign}

Allowing the radial interval to cross zero changes the root-counting argument: a negative root can supply the inner sign change that required an additional positive root in Appendix~\ref{app:pd-branch-proof}. We now determine which arrangements can make the local test fail under the assumptions of Section~\ref{sec:mixed-sign}, with $m > 0$. If $\kappa = 0$, the potential $\mathcal{R} = (r^2+c_\xi)^2$ has at most one positive root, so a larger positive root cannot block it. Every failure therefore has $\kappa > 0$ and
\begin{align}
    [r]\mathcal{R} = 2m\kappa>0.
\end{align}
To determine which root arrangements are compatible with this positive coefficient, consider a root $R > 0$ that passes the local derivative test but is blocked from the observer by a larger root. Let $S$ be the largest root below the observer. The endpoint signs require at least four roots (counted with multiplicity) by the argument in Appendix~\ref{app:pd-branch-proof}. At least three must be positive: those supplied by $R$, $S$, and either their multiplicities or an intervening root. The potential is therefore quartic, these roots account for the entire polynomial, and its positivity at the observer gives a positive leading coefficient $A$.

The fourth root must be strictly negative: if it were positive, Vieta's formula would give $[r]\mathcal{R} < 0$, and if it were zero, the product of the three positive roots would still give a negative linear coefficient. Both contradict $2m\kappa > 0$. The negative root is simple because the positive roots already account for the remaining three multiplicities. We can therefore write
\begin{align}
    \mathcal{R}(r) = A(r+u)(r-x)(r-y)(r-z),
\end{align}
with $u > 0$ and $0 < x \leq y \leq z$, allowing repeated factors. Vieta's relation then gives
\begin{align}
    2m\kappa = A\left[u(xy+xz+yz)-xyz\right]>0.
\end{align}
This proves
\begin{align}
    u > \frac{xyz}{xy+xz+yz}
\end{align}
and also
\begin{align}
    \mathcal{R}(0) = -A\,u x y z<0.
\end{align}
Since $\mathcal{R}(0) = c_\xi^2-\kappa \Delta_r(0)$ and $\kappa > 0$, it follows that $\Delta_r(0) > 0$, proving Equation~\ref{eq:mixed-zero-signs}.

There are three possible arrangements of the positive roots. If the hidden root $R$ is simple and the outer root $S$ has odd multiplicity, a sign-changing root is required between them. All three positive roots must then be simple to fit the quartic count, giving Pattern 1 of Section~\ref{sec:mixed-sign}. If $R$ is simple and $S$ has even multiplicity, the remaining two positive roots coincide at $S$, giving the asymptotic barrier of Pattern 2. Finally, a repeated hidden root can fit the quartic count only with multiplicity two and a simple larger root. The potential is negative on both sides of the hidden root, so $\mathcal{R}'' < 0$, giving Pattern 3.

Every other arrangement would require at least four positive roots (counted with multiplicity) and would give the same Vieta contradiction. Thus a hidden repeated root at positive radius must have $\mathcal{R}'' < 0$, and so an outer root cannot hide an unstable positive-radius orbit with $\mathcal{R}'' > 0$.

For one of these root patterns to occur on an observer's image, the ray must also follow a regular, angularly allowed path from the observer to the outer root. Pattern 1 occurs in Kerr--Newman--de Sitter with
\begin{align}
    m&=a=1, & \ell&=\alpha_{\rm acc}=0, \nonumber \\
    \Lambda&=\frac1{20}, & \beta_{\rm ch}&=\frac{43}{80}, \nonumber \\
    c_\xi&=\frac92, & \kappa&=60.
\end{align}
Then
\begin{align}
    \Delta_r = -\frac1{60}r^4+\frac{59}{60}r^2-2r+\frac{123}{80},
\end{align}
which is positive on $[-7,4]$, and
\begin{align}
    \mathcal{R}(r) = 2(r+6)(r-1)(r-2)(r-3).
\end{align}
At the equatorial observer $r_o = 4$, the constants correspond to the real strictly inward image point in adapted coordinates
\begin{align}
    X=-\frac98, \qquad Y=\pm\frac{\sqrt{159}}8.
\end{align}
The inward ray turns at $r = 3$, leaving the positive-slope root at $r = 1$ inaccessible. Since the ray stays within $3 \leq r \leq 4$, this failure of the local test occurs along a regular path. The extension through $r = 0$ contains the equatorial singularity $\Sigma(0,\pi/2) = 0$, while the NUT example in the main text also keeps $\Sigma$ nonzero across the entire $r = 0$ surface.

\section{Hidden photon orbits in the OP family}
\label{app:op-example}

The ray in Figure~\ref{fig:op-critical-counterexample} shares its constants of motion with an unstable photon orbit but turns before it can approach that orbit. In this case, the obstruction comes from two larger simple roots of the radial potential between the unstable orbit and the observer, all in the same regular exterior. In this Appendix, we specify the spacetime and photon constants, then verify that the ray's path is allowed both radially and angularly. We use the normalized Carter coordinates of the Ovcharenko--Podolsk\'{y} family and the separated photon equations derived in \cite{ovcharenko_2025_nonaligned,gray_2026_hidden}. After normalizing by the nonzero Killing energy, the radial and angular potentials are
\begin{align}
    \mathcal{R}(q) &= (q^2 + \xi)^2 - \kappa Q(q), \nonumber\\
    \Theta(p) &= \kappa P(p) - (p^2 - \xi)^2.
    \label{eq:op-separated-potentials}
\end{align}

\subsection{Metric functions and branch choice}

The OP field equations link the angular metric function $P$ to the radial function $Q$, so we can specify the angular geometry first and use it to determine both. We first choose the OP scale and angular coordinate so that $a = \omega = 1$, place the two polar axes at $y = \pm 1$, and write
\begin{align}
    y &= x - \lambda, \nonumber\\
    \overline{P}(y) &= (1 - y^2)(1 - A y + b y^2), \nonumber\\
    \mathcal{P}(x) &= \overline{P}(x-\lambda) = \sum_{j=0}^4 c_j x^j.
    \label{eq:op-normalized-angular}
\end{align}
The factor $1 - y^2$ fixes the axes, while $A$ and $b$ set the remaining coefficients of the angular quartic. The shift $\lambda$ and the combinations $s$ and $\tau$ are related to the OP parameters by
\begin{align}
    \lambda = \ell/a, \qquad s = \alpha C_{\rm OP}, \qquad \tau = \alpha a^2, \qquad C_{\rm OP} = \frac{s}{\tau},
\end{align}
The electromagnetic parameter $C_{\rm OP}$ belongs to the spacetime solution, while $\kappa$ remains the photon separation constant. To obtain the normalized mass $m$ and NUT parameter $n$ from the angular coefficients, we define
\begin{align}
    g & =1+s^2, & d &= 1-\lambda^2, \nonumber\\
    V &= \frac{gc_1}{2}, & U &= g\left(\frac{c_3}{2\tau} + s c_1\right), \nonumber\\
    m &= \frac{U-sV}{g}, & n &= \frac{V+sU}{g}.
    \label{eq:op-mass-nut-map}
\end{align}
Thus the same angular quartic that governs the polar motion fixes the mass and NUT parameters entering the radial function.

Solving the OP field equations requires choosing an algebraic branch. We use Branch I, which connects continuously to PD.\footnote{Ovcharenko and Podolsk\'{y} favor this branch for the black hole interpretation and question the physical relevance of the other branches they examine \cite{ovcharenko_2025_nonaligned}.} To retain the sign conditions that select Branch I, we define
\begin{align}
X_1 =& 1 - \tau^2C_{\rm OP}\left[2 + 6\lambda^2 - \tau^2C_{\rm OP}\,d^2 \right. \nonumber \\
&\left.\qquad + 8\tau^2C_{\rm OP}^2\lambda^2 + \frac{8\tau \,C_{\rm OP}\,d\,m\,\lambda}{c_0}\right], \nonumber \\
X_2 =& 1 - 4\tau^2C_{\rm OP} \left(1+\lambda^2 - \tau^2 \, C_{\rm OP}\,d^2\right), \nonumber \\
A_{\rm OP} =& 1 - \tau^2C_{\rm OP} \left(3 + 5 \lambda^2 - 2 \tau^2\,C_{\rm OP}\,d^2\right), \nonumber \\
F =& \frac{4\,\tau^4\,C_{\rm OP}^3\,d\,\lambda\, n}{g\,c_0} \nonumber\\
& + \frac{4\,\tau^3 \, C_{\rm OP}^2 \, d\, \lambda \,m}{g\,c_0}.
\label{eq:op-branch-data}
\end{align}
The Branch-I relation $A_{\rm OP} - F = \sqrt{X_1}\sqrt{X_2}$ can then be written without square roots as
\begin{align}
    (A_{\rm OP}-F)^2-X_1X_2&=0, \nonumber\\
    X_1&>0, & X_2&>0, & A_{\rm OP}-F&>0.
    \label{eq:op-branch-sheet}
\end{align}
The inequalities keep both expressions under the square roots positive and retain the positive sign of $A_{\rm OP}-F$, excluding the extra solutions introduced by squaring.

The radial shift $r_0$ is linked to an angular shift $x_0$. Once $r_0$ is chosen, the OP relation constrains $x_0$ to one of two values. To calculate them, we define
\begin{align}
    \epsilon &= -c_2,  \nonumber \\
    H &= (1+4s^2)\,c_0-2\,C_{\rm OP}\epsilon, \nonumber \\
    A_s &= \frac{2\,C_{\rm OP}\,V}{g\,H}, \nonumber \\
    B_s &= \frac{2\,C_{\rm OP}\,U}{g\,H}, \nonumber \\
    R_s^2 &= \frac{4\,C_{\rm OP}^2(U^2+V^2)}{g^2\,H^2}.
\end{align}
The example below uses the upper solution $z>0$ of
\begin{align}
    z^2=R_s^2-(r_0-B_s)^2, \qquad x_0=A_s+z.
    \label{eq:op-linked-shift}
\end{align}
In the centered Carter coordinates $p=x+x_0$ and $q=r+r_0$, the metric functions are
\begin{align}
    P(p)&=\mathcal{P}(p-x_0), &
    Q(q)&=\mathcal{Q}(q-r_0),
    \label{eq:op-centered-functions}
\end{align}
The radial quartic $\mathcal{Q}(r)=\sum_{j=0}^4B_jr^j$ follows directly from the angular coefficients:
\begin{align}
B_0&=c_0, & B_2&=-c_2, \nonumber\\
B_1&=-\left(\frac{c_3}{\tau}+2sc_1\right), \nonumber\\
B_3&=-2sc_3-\tau(1+4s^2)c_1, \nonumber\\
B_4&=c_4+\frac{2\tau s}{c_0}
\left[\left(\frac{c_3}{2\tau}+sc_1\right)^2
+\frac{c_1^2}{4}\right].
\label{eq:op-radial-partner-map}
\end{align}
Equations~\ref{eq:op-normalized-angular}--\ref{eq:op-radial-partner-map} therefore determine both metric functions from $(\lambda,A,b,s,\tau,r_0)$ and the chosen sign of $z$.

To verify that the photon path remains regular, we also need to check where the conformal factor can vanish. In the coordinates $x = p-x_0$ and $r = q-r_0$, the same parameters give
\begin{align}
\Omega^2 = {} & (1 - \tau rx)^2+2\tau^2C_{\rm OP}(r^2-x^2) \nonumber\\
              & + \tau^2\left(4\tau^2C_{\rm OP}^2 - \frac{2C_{\rm OP}\epsilon}{c_0}\right)r^2x^2 \nonumber\\
              & + \frac{4\tau^2C_{\rm OP}}{gc_0}(Ux+Vr)rx.
\label{eq:op-conformal-factor}
\end{align}
For each fixed angular coordinate $x$, this is a quadratic in $r$. On Branch I, its discriminant is
\begin{align}
    \mathrm{Disc}_r\,[\Omega^2]
    =-\frac{8s^2\mathcal{P}(x)}{C_{\rm OP}c_0}.
    \label{eq:op-conformal-discriminant}
\end{align}
Throughout the angular interval between the two axes, the example below has $\mathcal{P}(x)>0$, $C_{\rm OP}c_0>0$, and
\begin{align}
    \Omega^2(0,x) = 1-2 \,s\,\tau\, x^2 > 1 - 2\,s\,\tau \,(1+|\lambda|)^2 > 0.
    \label{eq:op-conformal-bound}
\end{align}
The negative discriminant excludes real zeros in $r$, and the value at $r = 0$ fixes the sign to be positive. Thus $\Omega^2 > 0$ throughout the angular interval for every finite $r$, including the full radial region used below.

We obtain a spacetime with these regularity properties and the required radial roots by fixing the rational parameters
\begin{align}
    (\lambda,b,s,\tau,r_0)
    =\left(-\frac{77}{80},\frac{1793}{10000},\frac{279}{2500},
    \frac{10591}{10000},\frac{1177}{10000}\right)
    \label{eq:op-witness-parameters}
\end{align}
Substituting these parameters into Equation~\ref{eq:op-branch-sheet} fixes the remaining angular coefficient $A$ as the unique root in
\begin{align}
    \frac{4235650853}{5000000000} < A <
    \frac{8471301707}{10000000000}
\end{align}
of the quadratic
\begin{align}
    0 =& 5912150769612959918894111099520000000000A^2 \nonumber \\
    & -8743147482900931376433837687989396800000A \nonumber \\
    & +3163850064633795500542970510190643519353.
    \label{eq:op-witness-A}
\end{align}
The root in this interval gives $A \simeq 0.8471301706$ and satisfies the Branch-I inequalities in Equation~\ref{eq:op-branch-sheet}. We take $z > 0$ in Equation~\ref{eq:op-linked-shift} to select the larger of the two angular shifts. For this spacetime, Sturm sign counts confirm that the mass is positive, $m \simeq 0.03430$, that $P > 0$ between the axes, and that $Q$ has a positive quartic coefficient. We retain the exact value of $A$ in the calculations below in order to ensure that the repeated-root conditions hold exactly.

\subsection{The unstable repeated orbit and its blockers}

In this spacetime, the ray shown in Figure~\ref{fig:op-critical-counterexample} has photon impact parameter
\begin{align}
    \xi = -\frac{427}{2000}.
\end{align}
The orbit radius must be a repeated root of the radial potential, with $\mathcal{R} = \mathcal{R}' = 0$. For $q^2 + \xi \neq 0$, eliminating $\kappa$ gives
\begin{align}
    K(q,A) \equiv 4\,q\,Q(q) - (q^2+\xi)\,Q'(q) = 0.
    \label{eq:op-stationary-equation}
\end{align}
We find the orbit radius by solving $K = 0$ together with Equation~\ref{eq:op-witness-A}. The quadratic relation for $A$ reduces the numerator of $K$ to an expression that is linear in $A$, and we solve for $A$ and substitute back into the quadratic, then clear denominators and remove the common integer factor to obtain a degree-eight polynomial in $q$. The orbit radius $q_\star$ is its unique root in
\begin{align}
    \frac{55274774576}{10^{11}} < q_\star < \frac{55274774577}{10^{11}}.
    \label{eq:op-repeated-root-bracket}
\end{align}
Exact polynomial reduction confirms that $q_\star$ satisfies the original equation $K = 0$ on Branch I. The remaining photon constant follows from requiring $\mathcal{R}(q_\star) = 0$, which gives
\begin{align}
    \kappa_\star = \frac{(q_\star^2 + \xi)^2}{Q(q_\star)}.
    \label{eq:op-critical-level}
\end{align}
Both repeated-root conditions then hold exactly:
\begin{align}
    \mathcal{R}(q_\star) = \mathcal{R}'(q_\star) = 0. \label{eq:op-blocked-repeated-root}
\end{align}
We use the following numerical bounds to verify the orbit's radial stability and the sign of $Q$ at its radius:
\begin{align}
    0.93801280942 &< \kappa_\star < 0.93801281005, \nonumber\\
    0.00902923049 &< Q(q_\star) < 0.00902923050, \nonumber\\
    0.04687534140 &< \mathcal{R}''(q_\star) < 0.04687534162.
    \label{eq:op-repeated-root-bounds}
\end{align}
The positive second derivative shows that $q_\star$ has multiplicity two and describes an unstable photon orbit, as required by Equation~\ref{eq:physical-fold-conditions}. The coefficient of $q^4$ in $\mathcal{R}$ lies between $0.0788554297$ and $0.0788554299$, and so radial motion is allowed beyond the largest root.

To find the turn reached by an observer's inward ray, we must locate the remaining radial roots. The potential has opposite signs at the endpoints of each of the following intervals:
\begin{align}
    \frac{17}{20} < q_3 < \frac{43}{50}, \nonumber \\
    \frac{77}{50} < q_4 < \frac{31}{20}.
    \label{eq:op-blocking-root-brackets}
\end{align}
Together with the root of multiplicity two at $q_\star$, these two roots account for all four roots of the potential, so both must be simple. The positive leading coefficient then gives $\mathcal{R} > 0$ for $q > q_4$ and $\mathcal{R} < 0$ between $q_3$ and $q_4$. An inward ray from $q_o = 20$ therefore turns at $q_4$, where the forbidden interval prevents it from reaching the unstable orbit. We also check that no horizon separates the observer from these photon radii. The four simple roots of $Q$, which locate the metric horizons, lie in
\begin{align}
    -1.16 &< h_1 < -1.15, \nonumber \\
    -0.07 &<h_2 < -0.06, \nonumber \\
    0.43 &< h_3 < 0.44, \nonumber \\
    0.47 &< h_4 < 0.48.
\end{align}
All three photon radii and the observer therefore lie beyond $h_4$ in the same connected exterior with $Q > 0$.

At large radius, $Q(q) \sim 0.982 \, q^4$, so the radial root curve $(q^2 + \xi)^2/Q(q)$ approaches a finite value. In asymptotically flat Kerr, the radial metric function grows quadratically and the root curve instead grows like $q^2$.

\subsection{Angular access and the finite-observer image point}

The photon must also have allowed angular motion along its path to the observer. We place the observer at $y_o = 559/850$, retaining the larger angular shift selected by $z > 0$. Equation~\ref{eq:op-centered-functions} gives $p_o = \lambda+y_o+x_0$ and $P(p_o) = \overline{P}(y_o)$. At this position, the angular coordinate and photon potential satisfy
\begin{align}
    -0.32168408828 &< p_o < -0.32168408819, \nonumber\\
    0.17656232097 &< \Theta(p_o) < 0.17656232122.
    \label{eq:op-angular-bound}
\end{align}
In this case, the photon can reach the observer's polar angle because $\Theta(p_o) > 0$. Since $P = 0$ and $\Theta < 0$ at both axes, its allowed angular motion is confined to an interval between two turning points away from the axes.

We locate the photon on the image using its momentum in the observer's local Carter frame:
\begin{align}
    X &= \frac{\xi-p_o^2}{\sqrt{P(p_o)}}, &
    Y &= \pm\frac{\sqrt{\Theta(p_o)}}{\sqrt{P(p_o)}}, &
    W &= \frac{q_o^2+\xi}{\sqrt{Q(q_o)}}.
    \label{eq:op-observer-screen}
\end{align}
A common factor $\Omega_o/\sqrt{p_o^2+q_o^2}$ has been removed from all three frame components normalized by $E$; this factor cancels from the sky coordinates $X/W$ and $Y/W$. The following identities can be used to connect this image point to the radial motion:
\begin{align}
    X^2 + Y^2 &= \kappa_\star, \nonumber\\
    \mathcal{R}(q_o) &= Q(q_o)\left(W^2-X^2-Y^2\right).
    \label{eq:op-screen-identities}
\end{align}
At the observer, $Q(q_o) > 0$ and $q_o^2 + \xi > 0$ give $W > 0$, so the photon is future directed. Since $\mathcal{R}(q_o) > 0$, its image point $(X/W, Y/W)$ lies strictly inside the unit disk, with
\begin{align}
    0.06593704764 < 1 - \frac{X^2+Y^2}{W^2} < 0.06593704842. \label{eq:op-screen-margin}
\end{align}
For the image point with $Y > 0$, we obtain
\begin{align}
    1.00211211182 &< W < 1.00211211191, \nonumber \\
    -0.58203664675 &< \frac{X}{W} < -0.58203664658, \nonumber \\
    0.77155446563 &< \frac{Y}{W} < 0.77155446626.
\end{align}
The photon moves toward increasing $q$ as it reaches the observer, so tracing it backward gives the inward ray shown in Figure~\ref{fig:op-critical-counterexample}. Its angular motion remains between the two turning points away from the axes. The conformal factor stays positive throughout this angular interval at every finite $q$, and $q \geq q_4 > 0$ keeps $q^2+p^2$ positive along the complete observer-to-turn path.

The repeated-root conditions in Equation~\ref{eq:op-blocked-repeated-root}, together with the positive second derivative, make this image point a candidate for the critical curve, but to determine whether it lies on the critical curve, we examine nearby rays. Since the outer root $q_4$ is simple, the implicit function theorem guarantees a nearby outer turn for every sufficiently small change of image point. The bounds above also keep the observer inside the allowed radial and angular regions and the path to the outer turn away from the axes, horizons, and singularities, so these properties persist for nearby rays. The outer root remains the first root reached and varies smoothly with image position. The selected point therefore lies inside the returning region, away from the physical critical curve.

The unstable orbit is therefore hidden by an outer turn within the same connected, regular exterior containing the observer, with $q > h_4 > 0$ and $Q > 0$.

\section{Images at the outer boundary}
\label{app:boundary-screens}

To extend the finite-observer construction to the outer boundary, we rescale the image coordinates as the observer approaches that boundary. In the standard asymptotically flat limit, $X$ and $Y$ decrease as $1/r_o$ for fixed photon constants. We can therefore define finite impact coordinates $(\bar X, \bar Y)$ by setting
\begin{align}
    X=\frac{\bar X}{r_o}, \qquad Y=\frac{\bar Y}{r_o}.
\end{align}
Every finite point $(\bar X, \bar Y)$ admits radial motion near infinity, since $\mathcal{R}(r) \sim r^4 > 0$ for fixed photon constants at large $r$. The distant image therefore covers the full impact plane. In the Kerr limit, these coordinates are related to Bardeen's convention by $\bar X = -\alpha-a \sin\theta_o$ and $\bar Y^2 = \beta^2$.

For nonaccelerating AdS with $\Delta_r \sim \mu r^4$, the same rescaling gives a bounded limiting screen. Its interior is the disk
\begin{align}
    \bar X^2+\bar Y^2 < \frac{1}{\mu},
\end{align}
equivalently $\kappa < 1/\mu$ in this normalization. This condition also makes the leading quartic term of the photon potential positive, allowing radial motion at sufficiently large radius.

With nonzero acceleration, the conformal boundary $\Omega = 0$ generally depends on angle, so we must specify where the observer approaches it. Near a regular boundary point at finite $r$, dividing $X$ and $Y$ by $\Omega_o$ removes their common vanishing factor and gives finite local image coordinates. To determine which rays appear in this limiting image, one must also specify the observer frame, the domain of outer communication, and the allowed angular interval. The rays must remain in the same connected region with $\Omega > 0$ as the observer.

\section{Algebraic properties of the conics and the critical curve}
\label{app:critical-curve-restrictions}

In Kerr and PD, the candidate critical curve satisfies a single polynomial equation in image coordinates. This appendix derives that equation from the radial potential and shows how its restrictions arise from requiring all conics to describe one spacetime and observer. We begin by identifying degenerate conics and giving a sufficient condition for repeated-root image points to vary smoothly with orbit radius.

\subsection{Conic degeneracy and repeated-root regularity}
\label{app:general-conic-details}

At special radii or parameter values, a candidate conic can become degenerate, just like what happens at the extremal Kerr horizon. To identify these cases, begin by collecting the coefficients of Equation~\ref{eq:general-root-conic} in a symmetric matrix $\mathsf Q_0(R)$. With $u = (\alpha, \beta, 1)^T$, the product $u^T \mathsf Q_0 u$ reproduces the fixed-radius conic equation:
\begin{align}
    \mathsf Q_0(R) =
    \begin{pmatrix}
        a_0&0&\ell_0/2 \\
        0&b_0&0 \\
        \ell_0/2&0&c_0
    \end{pmatrix},
    \qquad \mathcal{C}_0=u^T\mathsf Q_0u.
    \label{eq:general-conic-matrix}
\end{align}
The conic $\mathcal{C}_0 = 0$ is degenerate when the determinant of this matrix is zero,
\begin{align}
    \det\mathsf Q_0 = b_0\left(a_0c_0-\frac{\ell_0^2}{4}\right)=0. \label{eq:general-conic-degeneracy}
\end{align}
In particular, $b_0 = 0$ makes the equation independent of $\beta$ in the reflection-adapted chart. Affine or projective changes of image coordinates can introduce mixed and additional linear terms. The transformed curve is still a conic, and it is degenerate exactly when the original conic is degenerate.

To trace the critical curve by varying $R$, we also need the intersection of the root and derivative conics to vary smoothly. At a repeated-root candidate satisfying Equation~\ref{eq:physical-fold-conditions}, a sufficient condition is
\begin{align}
    \det \frac{\partial(\mathcal{C}_0,\mathcal{C}_1)}{\partial(\alpha,\beta)} \neq 0. \label{eq:repeated-root-transversality}
\end{align}
The determinant in Equation~\ref{eq:repeated-root-transversality} is built from derivatives of the two conic equations with respect to $\alpha$ and $\beta$. When it is nonzero, the conics cross with different tangent directions, and the implicit function theorem lets us follow their intersection smoothly as $R$ varies. Both conics are even in $\beta$ on the reflection-adapted image, so this determinant vanishes on the horizontal axis at $\beta = 0$. The image-plane curve can still be regular there, with a different local parameter taking the place of $R$. Tangential intersections, shared conic components, and higher-multiplicity radial roots require separate local treatments, as illustrated by the Kerr limits in Appendix~\ref{app:kerr-extensions}.

\subsection{Constructing ellipses tangent to a given curve}

To understand what constrains the shape of the critical curve, we first let the conic coefficients vary independently of a spacetime and observer. With this freedom, we can construct a family of ellipses whose envelope is any regular reflection-symmetric curve of the form
\begin{align}
    \Gamma_g =\{(X,Y):Y^2=g(X)\},
\end{align}
where $g \in C^2([x_-, x_+])$, $g \geq 0$, and $g + g'^2 > 0$. We choose the quadratic coefficient $L$ so that $g'' + 2L$ remains positive throughout the interval:
\begin{align}
    L>\frac12\lVert g''\rVert_\infty
\end{align}
For each $t \in [x_-, x_+]$, we define an ellipse $B(X, Y; t) = 0$, with
\begin{align}
    B(X,Y;t) =Y^2-g(t)-g'(t)(X-t)+L(X-t)^2. \label{eq:free-conic-envelope}
\end{align}
Varying $t$ gives a family of nondegenerate real ellipses. We find their envelope by differentiating $B$ with respect to $t$ at a fixed image point:
\begin{align}
    \partial_t B =-\left[g''(t) + 2L\right](X - t).
\end{align}
Since $g''+2L > 0$, the envelope conditions $B = \partial_t B = 0$ force $X = t$ and then $Y^2 = g(t)$, reproducing the entire prescribed curve $\Gamma_g$. The restrictions on Kerr and PD critical curves therefore arise from the relations among their conic coefficients, all of which must describe one spacetime and observer.

\subsection{The quartic radial discriminant}

For Kerr and PD, we obtain a single equation for the candidate critical curve by eliminating the orbit radius from the repeated-root conditions. The radial discriminant provides such a relation because it vanishes whenever the potential has a repeated root. To see this, write the PD-form potential as
\begin{align}
    \mathcal{R}(r) &=(r^2+c)^2-\kappa\Delta_r(r), & \Delta_r(r)&=\sum_{j=0}^4b_jr^j. \label{eq:critical-discriminant-radial}
\end{align}
At $\kappa = 0$, the potential is a perfect square for every $c$, so its discriminant vanishes. Evaluating the quartic discriminant gives the factorization
\begin{align}
    \mathrm{Disc}_r \, \mathcal{R} = \kappa^2 \mathcal{H}_\Delta(c, \kappa). \label{eq:reduced-radial-discriminant}
\end{align}
Dividing out the factor $\kappa^2$ defines the reduced discriminant $\mathcal{H}_\Delta$ and, unless special choices of spacetime parameters cancel the leading terms, the largest powers of $c$ and $\kappa$ are six and four. The powers of $c$ and $\kappa$ add to at most seven in any one term:
\begin{align}
    \mathrm{deg}_c \, \mathcal{H}_\Delta &= 6, & \mathrm{deg}_\kappa \, \mathcal{H}_\Delta &= 4, & \mathrm{deg}_{(c, \kappa)} \, \mathcal{H}_\Delta &= 7.
\end{align}
To write $\mathcal{H}_\Delta = 0$ in the image coordinates $(X, Y)$, we use the adapted GPL screen of Equation~\ref{eq:pd-screen-map} and define
\begin{align}
    c_o &= a^2 + \ell^2 - 2a\ell C - a\chi_o, & h &= a p_1.
\end{align}
Then
\begin{align}
    c=c_o-h X, \qquad \kappa=q_o(X^2+Y^2),
\end{align}
Substituting these expressions into Equation~\ref{eq:reduced-radial-discriminant} gives Equation~\ref{eq:pd-critical-octic}. For general spacetime and observer parameters, the resulting polynomial has total degree eight in $(X, Y)$, with only even powers of $Y$.

On the regular branch with exact multiplicity two, the radius parametrization is
\begin{align}
    c_c(R)
    &=\frac{4R\Delta_R}{\Delta_R'}-R^2,
    &
    \kappa_c(R)
    &=\frac{16R^2\Delta_R}{(\Delta_R')^2}.
    \label{eq:pd-critical-rational-parametrization}
\end{align}
It is valid away from $R \Delta_R \Delta_R' = 0$ and satisfies
\begin{align}
    \mathcal{H}_\Delta\left(c_c(R),\kappa_c(R)\right)=0.
\end{align}
Thus $c_c$ and $\kappa_c$ are ratios of polynomials in $R$, allowing us to trace the curve by varying the orbit radius. When $h \neq 0$, the image relations above give $X$ and $Y^2$ as ratios of polynomials in $R$ as well. Recovering the signed coordinate $Y$ generally requires taking a square root. For Kerr, the metric function $\Delta_r$ is quadratic, and removing the same universal factor recovers the sixth-degree image polynomial derived in Appendix C of \citet{gralla_2020_observableshape}. The Schwarzschild curve reduces to a circle, while extremal Kerr has both the exterior branch and the separate horizon component derived in Appendix~\ref{app:kerr-extensions}. Vanishing rotation, face-on charts, repeated horizons, and drops in polynomial degree likewise require their limiting equations.

When the quartic term vanishes, the photon potential becomes a cubic or takes on an even lower order. The quartic discriminant treats the missing finite roots as roots at infinity, so it can vanish even when the remaining finite roots are simple. In these limits, $\mathcal{H}_\Delta = 0$ still contains every finite repeated-root point with $\kappa \neq 0$, but it may also contain components produced by the degree reduction.

Removing the $\kappa^2$ factor preserves every repeated-root point with $\kappa \neq 0$ and also leaves the physical positive-radius critical curve intact, because the discarded branch has no positive radial root on the real GPL screen. There, $\kappa = 0$ requires $X = Y = 0$ and
\begin{align}
    c=c_o=(\ell+a\cos\vartheta_o)^2\geq0,
\end{align}
so $\mathcal{R} = (r^2+c_o)^2$ has no positive-radius root. The physical critical curve is selected from the remaining algebraic curve by retaining real image points with positive local energy whose rays can approach a repeated root with $\mathcal{R}''(R) > 0$ along a regular, angularly allowed path. These propagation conditions can select one or more finite arcs of the same algebraic curve.

\bibliography{references}

\end{document}